\documentclass{aa}  
\usepackage{graphicx}
\usepackage{txfonts}
\begin{document}
%

\title{Hydrogen and Helium traces in Type Ib-c Supernovae}
\titlerunning{CCSNe spectra}  

  \author{A. Elmhamdi \inst{1,2,3}, I.J. Danziger \inst{3}, D. Branch \inst{4}, B. Leibundgut \inst{5}, 
   E. Baron \inst{4}, R.P. Kirshner \inst{6}}

   \offprints{A. Elmhamdi, { elmhamdi@ts.astro.it;  elmhamdi@sissa.it}}
   
   \institute{ICTP-International Center for Theoretical Physics, Strada Costiera 11, 34014 
         Trieste - Italy
         \and INAF-Osservatorio Astronomico di Callurania, Via Mentore Maggini, 64100 
         Teramo - Italy
	 \and INAF-Osservatorio Astronomico di Trieste, Via G.B.Tiepolo 11 - I-34131
	 Trieste - Italy 
    	 \and Department of Physics and Astronomy, University of
	 Oklahoma, Norman, OK 73019 - USA
	 \and European Southern Observatory, Karl-Schwarzschild-Strasse 2. D-85478, Garching - 
          Germany
         \and Harvard-Smithsonian Center for Astrophysics, 60 Garden
	 Street, Cambridge, MA 02138 - USA}
 
\authorrunning{Elmhamdi et al.}

\date{}
\abstract
{}
{To investigate the spectroscopic properties of a selected optical photospheric 
 spectra of core collapse supernovae (CCSNe). Special attention is devoted to 
 traces of hydrogen at early phases.
 The impact on the physics and nature of their progenitors is emphasized.}
{The CCSNe-sample spectra are analyzed
 with the parameterized supernova synthetic spectrum code ``SYNOW'' adopting some
 simplifying approximations.}
{The generated spectra are found to match the observed ones
reasonably well, including a list of only 23 candidate ions. Guided by SN Ib 1990I,
 the observed trough near 6300\AA~is attributed to H$\alpha$
in almost all Type Ib events, although in some objects it becomes too weak to be 
 discernible, especially at later phases. Alternative line identifications are discussed.
 Differences in the way hydrogen manifests its presence within CCSNe are highlighted. 
 In Type Ib SNe, the H$\alpha$
contrast velocity (i.e. line velocity minus the photospheric
velocity) seems to increase with time at early epochs,
reaching values as high as 8000 km s$^{-1}$ around $15-20$ days after
maximum and then remains almost constant.
 The derived photospheric velocities, indicate a lower velocity for Type II SNe 1987A
and 1999em as compared to SN Ic 1994I and SN IIb 1993J, while Type Ib events
display a somewhat larger variation. 
The scatter, around day 20, is measured to be $\sim$5000 km s$^{-1}$.
Following two simple approaches, rough estimates of ejecta and 
 hydrogen masses are given.
A mass of hydrogen of approximately  0.02 $M_\odot$ 
is obtained for SN 1990I, while SNe 1983N and 2000H ejected $\sim$0.008 $M_\odot$ and
$\sim$0.08 $M_\odot$ of hydrogen, respectively. SN 1993J has a higher hydrogen mass,
$\sim 0.7$ $M_\odot$ with a large uncertainty. A low mass and thin hydrogen layer with very high
ejection velocities above the helium shell, is thus the most likely scenario
for Type Ib SNe. Some interesting and curious issues relating to
 oxygen lines suggest future investigations.}
{}
\maketitle
\section{Introduction}
Stripped-envelope SNe, namely Type Ib (helium-rich) 
and Type Ic (helium-poor), being hydrogen deficient objects, are undoubtedly amongst the most mysterious  SN 
classes. Recently efforts have started to understand the nature of these objects through studies of samples 
(Matheson et al. 2001\nocite{Math01}; Branch et al. 2002\nocite{Br02}). 

However the rarity of cases with well sampled observations, photometry and spectra, hampers a more direct 
inference of the physical situation behind the explosions and hence a clear view of the progenitor nature.
 Nevertheless, the discovery of metamorphosing events as SN 1987K and SN 1993J (recognized as SNe IIb), that 
evolve from 
Type II to Type Ib-c as they age, together with the similarity of the environments in which they occur have 
linked Type Ib-c SNe to a core-collapse scenario in massive stars.
At present, indeed, the most widely accepted models relate Type Ib-c SNe to both relatively low mass 
progenitors within
the context of close binary system evolution (i.e. mass-loss as consequence of mass transfer) and massive stars 
that have undergone significant mass-loss due to a wind (i.e. Wolf-Rayet stars). So far observations have not 
discriminated between the two scenarios, although recently, using HST data, a high spatial resolution
 search for the progenitor of the Type Ic SN 2004gt in a wide wavelength 
 range from the far UV to the near IR
 has suggested that the event might result from an evolved Wolf-Rayet star, although the 
 observations could not constrain models invoking less massive progenitors in binary systems (Gal-Yam et 
 al. 2005). 

Photometrically, the  lack of significant hydrogen in the outer layers
of SNe~Ib-c probably inhibits the most 
important characteristic of Type II SNe light curves, namely the plateau phase resulting from the hydrogen 
recombination wave.

At late phases, the steeper decline rate, compared to the ``$^{56}$Co to $^{56}$Fe'' decay slope, is indicative 
of significant $\gamma$-ray escape as a result of the low mass ejecta in this class of objects 
(Clocchiatti $\&$ Wheeler 1997\nocite{Cloc97}); there are rare exceptions where the late slope approaches the 
full trapping rate (e.g. SN Ib 1984L; Schlegel $\&$ Kirshner 1989\nocite{Schl89}).

Spectroscopically, a clear separation scheme within the
stripped-envelope SNe subclasses is still lacking.
 Part of the problem is the absence of meaningful statistics. A direct classification was earlier proposed by
 Harkness et al. (1987)\nocite{Har87} on the basis of He I strengths in the photospheric
 optical spectra of Type Ib SNe. Wheeler et al. (1994\nocite{Whee94}), however, have claimed the presence  of He I 10830\AA~
 in SN Ic 1990W, although He I lines were not noted in the optical region. The authors presented the idea of 
adopting, instead, the OI 7773\AA~ line as a distinguishing feature. The
absorption seems stronger in Type Ic than in Ib SNe. Matheson et al. (2001)\nocite{Math01} came to the same conclusion when 
analyzing a sample of Ib and Ic events. 
In addition, it has been argued that in Type Ib SNe 
He I lines 5876\AA~ and 7065\AA~ gradually grow in strength with respect to the one at 6678\AA~(Matheson et 
al. 2001)\nocite{Math01}. SN 1998bw (GRB980425) is another example of classification as Type Ic, but where IR lines of 
He I have been clearly identified (Patat et al. 2001)\nocite{Pat01}.

Branch et al.(2002)\nocite{Br02} have presented a relation between the velocity at the photosphere, measured using 
 synthetic-spectrum analysis, and the time since maximum light for a sample of Type Ib SNe. The relation is well 
 fitted by a power-law,
 indicating a fairly homogeneous behaviour. Interestingly, SN Ib 1990I does not follow this trend 
(Elmhamdi et al. 2004)\nocite{Elm04}.

 An important issue concerns features in the 6000$-$6500\AA~region of the early spectra in Ib-c SNe. Deng et 
 al. (2000)\nocite{Den00}, when analyzing spectra of SN Ib 1999dn, have attributed the absorption feature seen at 6300\AA~
 around maximum brightness to H$\alpha$  which later on disappears or is overwhelmed by C II 6580\AA~when 
 H$\alpha$ optical depth decreases as a consequence of the envelope expansion. The possibility of the presence 
 of C II 6580\AA~in Ib spectra has been earlier suggested by Harkness et al.(1987)\nocite{Har87}.
 Alternatively, the 6300\AA~absorption was identified to be due to Ne I 6402\AA~in SN Ib 1991D (Benetti et al. 
 2002). Ne I lines and H$\alpha$ had already been proposed to account for the deep absorption in SN Ib 1954A 
 by Branch (1972)\nocite{Br72}.
 
 The presence, or not, of hydrogen and/or helium in Type Ib-c, with the possibility of quantifying the amount, 
 is of great importance in identifying the progenitor stars that may give rise to these classes of objects.  
 
 The papers by Matheson et al. (2001)\nocite{Math01} and Branch et al. (2002)\nocite{Br02} have provided
 an impetus towards an advanced understanding of SNe Ib and Ic. The present work may be regarded as a 
 continuation of those efforts. However in our comparative study of early spectra of Type Ib SNe we include 
 representatives of all the various types of CCSNe, namely Type IIb, Type Ic and Type II. This is established 
 by means of synthetic spectra generated with the parametrized SN synthetic-spectrum code ``SYNOW''.

  Our main goal was to understand the similarities and differences among SNe Ib objects in the available sample,
  and to also compare to properties of the wider CCSNe family. 
 The paper is organized as follows. First the analysis method 
 and parameters are illustrated in Sect. 2. Data description is briefly given and the best fit synthetic 
 spectra are presented and compared with the observed ones in Sect. 3. This is done 
 separately for each individual object among  representatives of CCSNe classes. Sect. 4 presents two methods  
 to obtain spectroscopic mass estimates. The complete results will be discussed in detail and conclusions will 
 be drawn in Section 5.            
\section{Fitting procedure: SYNOW code} 
For the purpose of our analysis we make use of the parameterized supernova synthetic spectrum code ``SYNOW''. The code 
assumes spherical symmetry, homologous expansion: ``v $\propto$ r'' velocity-law and resonant scattering line formation 
above a sharp photosphere, emitting a blackbody continuum. 
SYNOW treats line formation and line blending (i.e. multiple scattering) within the Sobolev 
approximation (Jeffery \& Branch 1990\nocite{Jeff90}; Fisher 2000\nocite{Fish00}; Branch 2001\nocite{Br01}), and reads 
 lines from the Kurucz 42-million atomic 
line list (Kurucz 1993)\nocite{Kur93}. It is an LTE code in only one respect: LTE excitation 
 for the relative strengths of the lines 
of an individual ion, but it does not assume LTE ionization.

SYNOW provides a number of free fitting parameters. The most important are: 1. ($\tau _{ref}$), the optical depth 
of the strongest line,  in the optical region, of the introduced ion. The line is 
 called ``the reference line''. The optical depths of the other lines of the same ion are thereafter determined 
assuming Boltzmann equilibrium. To decide which ion to introduce in the synthesis procedure we rely on 
the work by Hatano et al.(1999)\nocite{Hat99} as a starting point. The authors have presented the variation 
 with temperature of 
LTE$-$Sobolev line optical depths of 45 individual candidate ions that might 
be encountered in supernova envelopes for six different compositions. 
2. ($T_{bb}$): the underlying blackbody continuum temperature. We did not attach high physical 
 importance to this parameter
 although the galactic extinctions for each individual supernova, as reported by Schlegel et al. 1998, is taken
 into account.  Independently  of the present study, estimating the total reddening to the event, both that caused by 
 foreground dust in the Milky Way and that caused by dust in the host galaxy, is a crucial point in supernovae
 study, especially when using them as cosmological probes. 
3. ($v_{phot}$): the velocity at the photosphere, estimated  from 
the match with Fe II lines. Restrictions on the velocity interval within which an ion is introduced is possible. A 
maximum outer boundary velocity of line-forming-region of $5 \times 10^{4}$ km s$^{-1}$ is adopted in the present 
analysis. 
When assigning a minimum ion velocity, ``$v_{min}$'', greater than
``$v_{phot}$'', the ion is said to be ``detached'' from the photosphere, and consequently has a ``non-zero'' 
optical depth only starting at ``$v_{min}$''. 
In SYNOW the profile of a detached line has a flat-topped
emission, and the absorption minimum is blueshifted by the detachment
velocity. An undetached line has a rounded emission peak. A slightly
detached line has a flat top but only over a small wavelength interval.
However whenever one talks about detached lines one needs to keep in mind that we are using a somewhat unrealistic 
``$\tau_v$'', one that has a discontinuity in it. Real supernovae spectra probably do not have sharp discontinuities. 
    
The radial dependence of the line optical depths can be chosen to be either exponential with an $e$-folding 
velocity ``$v_e$'' as free 
parameter (i.e. $\tau \propto exp(-v/v_e)$), or a power-law with an index ``$n$'' (i.e. $\tau \propto v^{-n}$). In 
 the present work a power-law profile is adopted with an index $n=8$, although for some objects we will discuss the
  possibility of an exponential profile.

 When fitting our CCSNe sample spectra we have tested many combinations of fitting parameters. Only the best generated 
 synthetic spectra are displayed in the following and compared to the observed ones.
 We introduce the parameter ``contrast velocity'', defined as the line minimum velocity minus to 
 the photospheric one: ``$v_{cont}(line)=v_{min}(line)-v_{phot}$''. We show as well the behaviour of a similar 
 parameter, defined instead as a ratio, i.e. ``$v_{cont}^{ratio}(line)=v_{min}(line)/v_{phot}$''.
   
 Although the computations are made under the purely resonant scattering assumption, we find
 that the photospheric-optical spectra of CCSNe are fitted well usually requiring only 23 candidate ions
 or fewer. Table 1 lists the candidate reference lines sufficent to reproduce the observed features in 
  optical spectra of the CCSNe sample. It is important to note that 
 since at early phases line formation takes place in high velocity layers causing severe line blending, it is better 
 in some complicated cases to analyze line identifications in reverse chronological order (i.e. starting with later 
 phase spectra followed by earlier ones). All the analyzed spectra have been transformed to the rest frame of their 
 host galaxies.  
\begin{table}
\begin{minipage}{80mm}
\caption{The candidate reference lines of CCSNe phtospheric-phase spectra in the optical 
 region (shown in increasing wavelength order).} 
\centering
\begin{tabular}{ccc}\\
\hline \hline
$~~~~$Ion  & Rest Wavelength (\AA)\\  
\hline
$~~~~$Ca II &3934 \\
$~~~~$Ni II &4067 \\
$~~~~$Co II & 4161 \\
$~~~~$Mn II& 4205 \\
$~~~~$Ca I & 4227 \\
$~~~~$Cr II & 4242 \\
$~~~~$Sc II &4247 \\
$~~~~$C II & 4267 \\
$~~~~$Mg II & 4481 \\
$~~~~$Ti II & 4550 \\
$~~~~$Ba II & 4554 \\
$~~~~$Fe II & 5018 \\
$~~~~$Mg I & 5184 \\
$~~~~$He I & 5876 \\
$~~~~$Na I & 5890 \\
$~~~~$Si II & 6347 \\
$~~~~$Ne I & 6402 \\
$~~~~$H I & 6563 \\
$~~~~$$[$O II$]$ & 7321 \\
$~~~~$O I & 7772 \\
$~~~~$Si I & 7944 \\
$~~~~$N I & 8680 \\
$~~~~$C I & 9095 \\
\hline \hline
\end{tabular}
\end{minipage}\\
\normalsize
\end{table}
\section{Data description and Analysis}
Our selected CCSNe sample consists of 20 objects - 16 of them are Type Ib, 2 are Type Ic, 1 is Type IIb and 1 is 
Type II SN. Some of the spectra presented in this work were gathered with the 60"
Telescope and the ``MMT'' on Mount Hopkins. They are part of the supernova
monitoring program of the Center for Astrophysics (PI R.P. Kirshner). Data for SN 1996aq are 
 taken from the Padova-Asiago supernova database. Use is made as well of the 
 ``SUSPECT''\footnote{http://bruford.nhn.ou.edu/~suspect/index1.html} supernovae spectral archive.
Descriptive data regarding the sample events are listed in Table 2 (i.e. host galaxy, recession velocity 
and phases). For each individual event, the table summarizes as well the most important fitting parameters. The
 last column indicates the number of ions we find responsible for determining the best fit spectra
 (details described in the following).
 Throughout the present work, however, we do not provide details concerning observations of individual 
objects. We focus, instead, on synthetic spectra 
 fitting procedures, constraints, problems and what we may learn about the CCSNe physical situation.   
\subsection{Type Ib SNe: the sample}

$*$ {\bf{SN 1990I:}}

We start with SN 1990I since it can be considered as one of the better observed objects among Type Ib-c SNe in terms 
of quality and sampling.  The event has been exploited both photometrically and spectroscopically giving constraints 
on ejecta, oxygen and nickel masses and energy estimates (Elmhamdi et al. 2004). The supernova seems to follow a 
velocity trend different from the pattern shown by the 10 SNe Ib sample of Branch et al. (2002). Here we investigate 
this peculiarity by means of synthetic spectra fits. 

The observed spectrum at maximum light, shown in Fig. 1, is compared to a synthetic spectrum (SSp) that has a velocity
 at the photosphere $v_{phot} = 12000$ km s$^{-1}$ and a blackbody continuum 
temperature T$_{bb}=$14000 K. The SSp contains lines of He I, Fe II, Sc II, Mg II, O I, Ca II and H$\alpha$. The lines
 of Sc II, with $\tau$(Sc II)$\sim 0.4$, have been introduced to help the fit of the absorption feature around 6000\AA~
and redward the trough attributed to He I 5876\AA. Absorption
troughs of P-Cygni He I lines at 5876\AA, 6678\AA~ and 7056\AA~are evident, although their relative strengths cannot 
 be simultaneously fitted  within the LTE
 approach in the SSp. We will face this limitation each time we analyze and fit He I lines. A more precise analysis  
 requires NLTE treatment as He I lines may be
 non-thermally excited by the decay products of $^{56}$Ni and $^{56}$Co (Lucy 1991)\nocite{Lucy91}. 
 Apart from He I and H$\alpha$ lines, the remaining lines have non-zero optical depths starting at the photospheric 
 velocity (i.e. they are undetached). The SYNOW parameters required to account for the He I lines are 
 $\tau$(He I)$\sim 2.9$ and $v_{min}$(He I)$=14000$ km s$^{-1}$. 

\begin{figure}
\includegraphics[height=9cm,width=9cm]{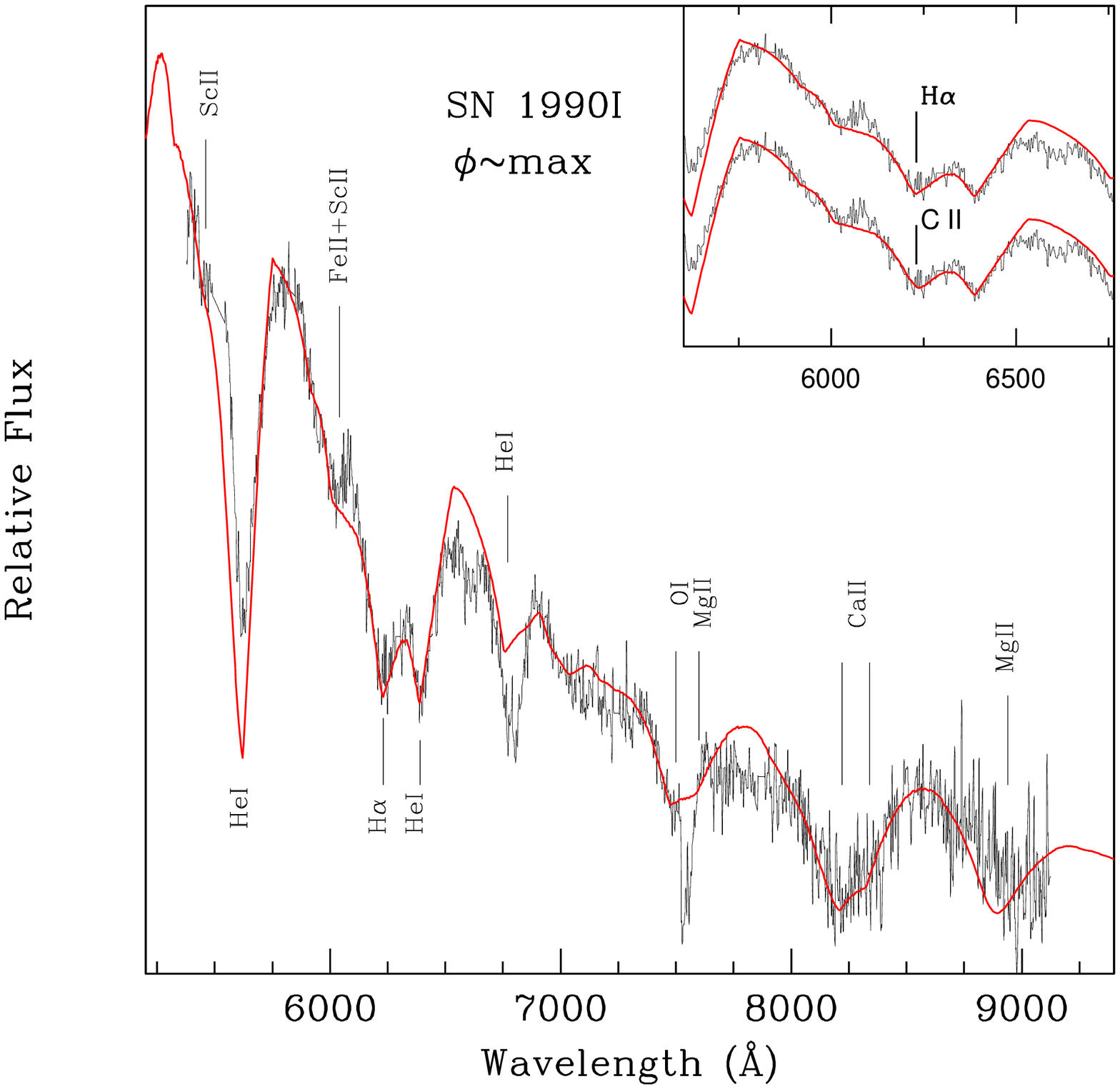}
\caption{ The SN 1990I spectrum, near maximum, compared with the ``SYNOW'' SSp (thin line). Lines that 
 are responsible for the most conspicuous features are also shown. The region around the 6250\AA
 ~trough is zoomed in the window.}
\end{figure}                                      
The absorption minimum near 6250\AA~is well fitted by H$\alpha$, with $v_{min}$(H$\alpha$)=16000 km s$^{-1}$ and 
assigned a moderate optical depth of 0.6 (Fig. 1).  However, we tested other plausible alternative identifications 
 that were introduced in the literature to account for similar features seen in Type Ib SNe, namely Si II 6355\AA, 
 Ne I 6402\AA~and C II 6580\AA~lines.
 Undetached Ne I lines were quite tempting in SN Ib 1991D, and worked as well
as H$\alpha$ having a contrast velocity of $v_{cont}$(H$\alpha$)
=7000 km s$^{-1}$ (Benetti et al. 2002)\nocite{Ben02}. For SN 1990I, instead, $v_{cont}$( H$\alpha$) is only 
 4000~km~s${-1}$ which means that 
 undetached Ne I line is excluded because it is  too blue to fit the observed feature. Si II 6355\AA~is obviously also 
too blue. C II 6580\AA~remains then a plausible alternative for H$\alpha$, since its rest wavelength is slightly redder
 than  H$\alpha$ and therefore has its contrast velocity as a free parameter. In the window of Fig. 1, we show 
synthetic spectra for both cases (H$\alpha$ and C II). C II, with $v_{min}$(CII)=17000 km s$^{-1}$ and
 $\tau$(CII)$\sim 0.003$, provides a fit as good as H$\alpha$. But an acceptable C II 6580\AA~line would require an 
 additional velocity of about 820 km s$^{-1}$  compared to  H$\alpha$ ($\sim 18$\AA~difference). C II 6580\AA~has been 
 proposed in early spectra of SN Ib 1999dn (Deng et al. 2000). Indeed, at maximum light and day 14, SSp fits of SN 
 1999dn included C II with  $v_{cont}$(CII)$=$ 7000 and 1000 km s$^{-1}$ respectively, while He I lines were formed 
 starting at the photosphere (i.e.$v_{cont}$(He I)$=$0 km s$^{-1}$) (Deng et al. 2000).
   For SN 1990I, the situation is almost similar: He I in the SSp shown in Fig. 1 has 
 $v_{cont}$(He I)$=$2000 km s$^{-1}$ while $v_{cont}$(CII)$=$5000 km s$^{-1}$. This means that the C II lines would 
 need to be formed above the helium layer, which would be surprising, if not unphysical, in Type Ib SNe. 

 SN 1990I, with its particular velocity structure, presents good evidence
  for an H$\alpha$ feature. However since it is not absolutely clearcut that H$\alpha$ is always responsible for the 
 absorption at about 6300\AA~seen in Type Ib SNe, the presence of the H$\beta$ Balmer line would be of great support. 
 Unfortunately the optical depth sufficient to fit the H$\alpha$ trough is so small that the other Balmer features 
 are too weak to be unambiguously detectable ($\tau$(H$\alpha)$ is about 7 times greater than $\tau$(H$\beta)$).  
\begin{figure}
\includegraphics[height=9cm,width=9cm]{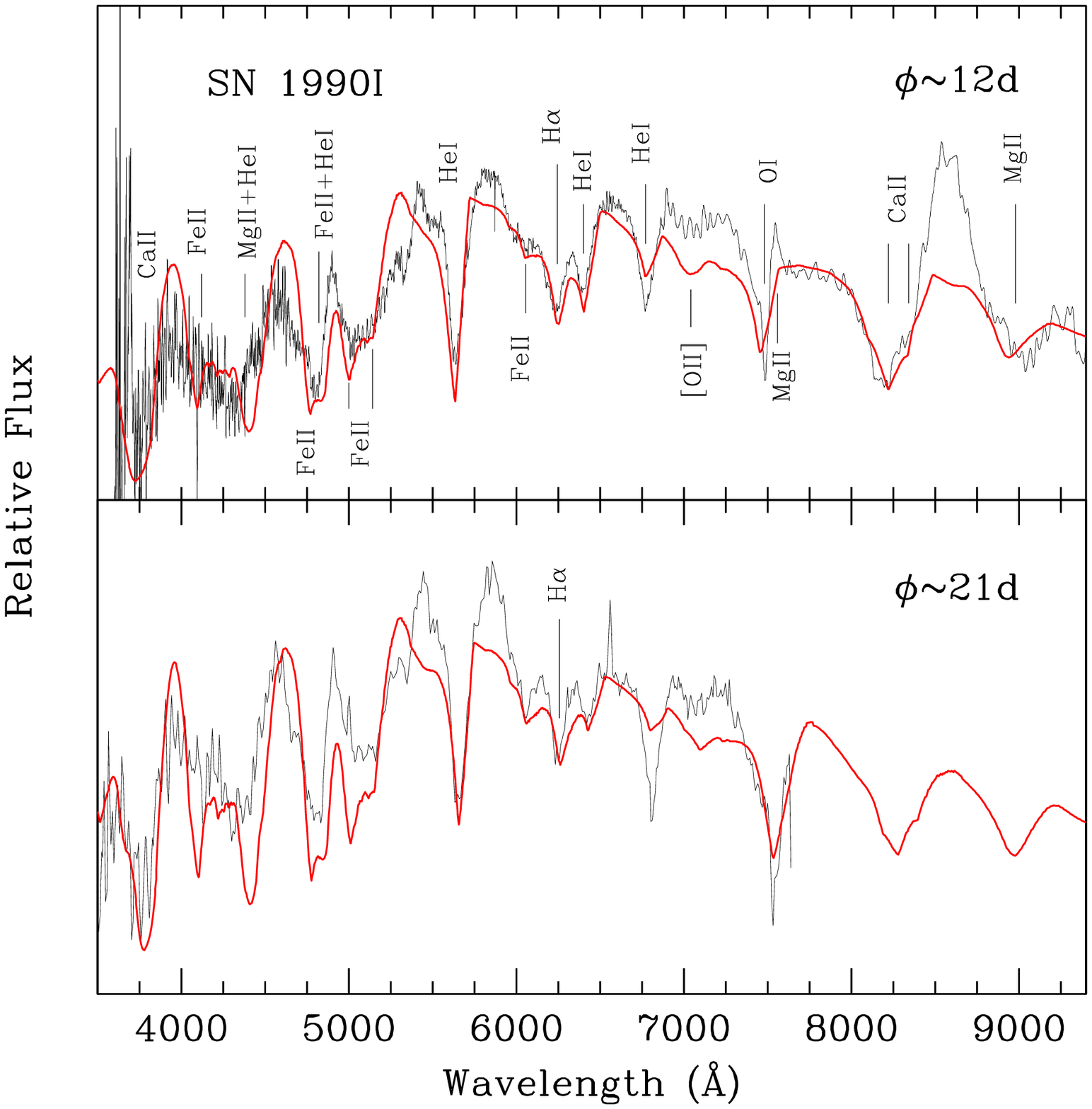}
\caption{SN 1990I ``SYNOW'' fits of the observed 12day and 21day spectra. Conspicuous line features are 
 indicated}
\end{figure}                                      

 Based on our investigation, the ions that generally might be encountered in shaping the  6000$-$6500\AA~wavelength 
 range are:  H$\alpha$, Ne I, C II, Si II, Sc II, Ca I, He I, Fe II, Si I and Ba II. Here, we propose the following 
 criteria in view of the methodology for deciding what lines should be adopted 
 (especially for the 6300\AA~feature):\\ 
  $\bf{1/-}$ when Fe II lines are very strong, they could produce a trough that might be sufficient to fit the 
 6300\AA~feature. The depth of that Fe II feature is controlled by means of Fe II lines at  4924, 5018 and 5169\AA.\\  
 $\bf{2/-}$ undetached Sc II, Si I, Ca I and Ba II can be introduced especially to fit the slope at 6000$-$6300\AA~
 wavelength range. They should not, however, introduce unwanted features in the rest of the spectrum.\\
 $\bf{3/-}$ undetached Ne I 6402\AA~line is rejected once its feature is too blue to fit the 6300\AA~trough or/and 
 when the other Ne I lines clearly introduce various unwanted features. Similar reasoning applies to the Si II 6355\AA~
 line.\\
 $\bf{4/-}$ with its contrast velocity as a free parameter, C II 6580\AA~line could be a candidate for the 6300\AA~
 trough; nevertheless it is ruled out once it exceeds the He I contrast velocity.
         
\begin{figure}
\vspace{0.3cm}
\includegraphics[height=7.5cm,width=8cm]{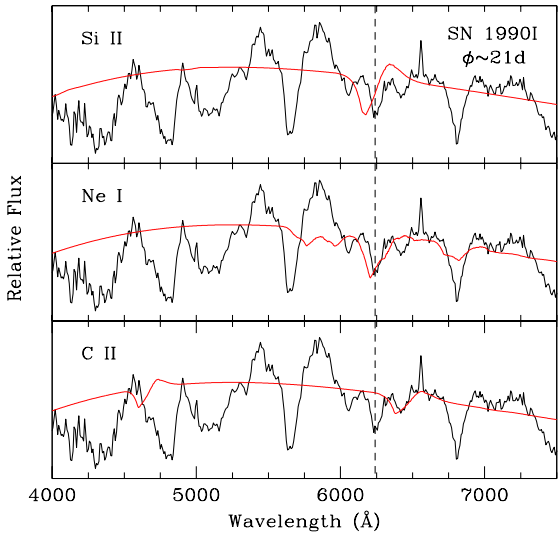}
\caption{ Synthetic ``SYNOW'' fit of the 21days observed spectrum, with only lines of SiII (top), NeI 
 (middle) and CII (bottom).}
\end{figure}                                      

 In Figure 2, the observed spectra at 12days (top panel) and 21days (bottom panel) are compared to the synthetic ones 
 with ($v_{phot}$=10000 km s$^{-1}$; T$_{bb}$=5500 K) and ($v_{phot}$=9500 km s$^{-1}$; T$_{bb}$=5400 K), respectively.
 Synthetic line profile features are labeled by the designation of the ion whose line gives rise to the feature. He I
 is still detached from the photosphere with $v_{cont}$(He I)$=$ 3000 km s$^{-1}$ at 12d and  $v_{cont}$(He I)$=$ 2500
 km s$^{-1}$ at 21days. The He I 7065\AA~seems to increase in strength relative 
 to the He I lines at 5876\AA~ and 6678\AA, indicating the non-thermal excitation effects are changing but still 
 existent. Two interesting points emerge from our SSp fit: on the one hand, even though the reference line Ca II 
 3933\AA, with $\tau$(Ca II)=120 at both 12days and 21days, produces a good match to the observed one, the 
 Ca II infrared triplet (8542, 8662, 8498\AA) has a clear deficit especially in the emission component of its P-Cygni 
 profile. On the other hand, the observed O I 7773\AA~is deep. An optical depth of 2 is in fact imposed to reproduce 
the deep trough in the SSp at 21days. The O I 7773\AA~lines are believed to have relatively greater strength in Type Ic
 SNe compared to Ib objects (Wheeler et al. 1994\nocite{Whee94}; Matheson et al. 2001). This is presumably 
 because for a ``$bare$'' 
 C/O envelope of Type Ic, one would expect oxygen lines to be more prominent relative to Ib case where an intact helium,
  and possibly some hydrogen, could tend to dilute the C/O core.

 Matheson et al. (2001) have defined a parameter called ``Fractional Line Depth'' of the line through absorption 
 minimum relative to the continuum flux. 
 Mean values of 0.27 ($\pm$0.11) and 0.38 ($\pm$0.091) were
 found to represent Type Ib and Type Ic respectively. For the case of SN 1990I, we measure a value of about 0.45, 
 indicating that the SN was peculiar in this respect. One explanation for this abnormal behaviour of SN 1990I might 
 be related to the amount of oxygen. In fact Elmhamdi et al. (2004) have argued for a possible high oxygen mass 
 ($\sim$0.7$-$1.35 M$_\odot$) relative to other Ib objects  ($\sim$0.3 M$_\odot$ in SNe 1984L, 1985F and 1996N).      
\begin{figure}
\includegraphics[height=9cm,width=9cm]{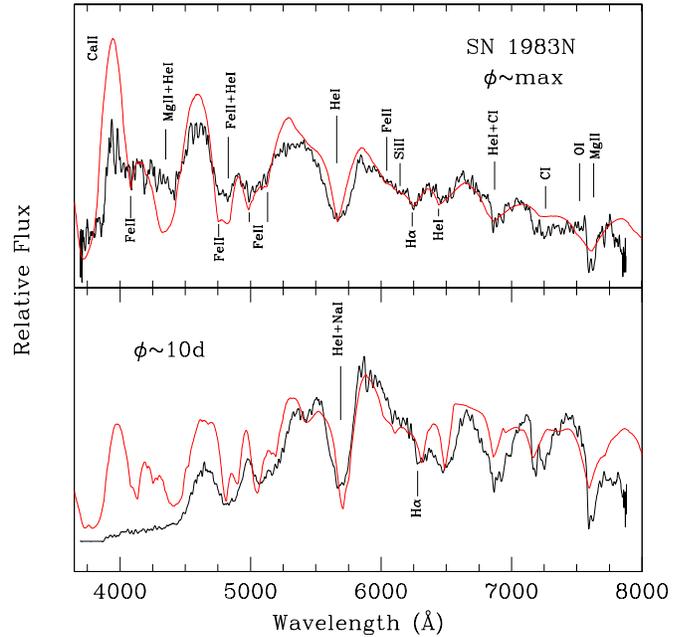}
\caption{ SSp fit of SN 1983N compared to the observed spectra at maximum light (upper panel) and
 at 10days (lower panel). Conspicuous line features are shown.}
\end{figure}                                      

 As far as the 6000$-$6500\AA~wavelength region is concerned, we find that Fe II, He I and H$\alpha$ are sufficient to 
 reproduce the overall shape in the SSp at both 12days and 21days. We have, however, tested the Ne I, Si II and C II 
 possibilities as candidates for the 6250\AA~trough. Panels in Fig. 3 are similar to the bottom of Fig. 2, but with 
 only Si II(top), Ne I(middle) and C II(bottom) lines. All three ions are formed starting at the photosphere 
 ($v_{cont}$= 0 km s$^{-1}$). Both Si II and Ne I lines are ruled out by means of criterion number 3 above. C II is 
 as well rejected because of criterion 4. We therefore consider H$\alpha$ to be the most likely explanation in SN 1990I.
\vspace{0.4cm}\\
$*$ {\bf{SN 1983N:}}

The observed spectra at maximum light and at $\sim$10days are shown in Fig. 4, together with the corresponding 
 synthetic spectra. The SSp at maximum brightness has  $v_{phot} = 11000$ km s$^{-1}$ and a blackbody continuum 
 temperature T$_{bb}=$8000 K, while the one at 10days has $v_{phot} = 7000$ km s$^{-1}$ and T$_{bb}=$5000 K. Ne I 
 and C II are rejected because of the criteria 3 and 4, while undetached Si II helps in fitting the absorption
 blueward  the 6250\AA~trough. H$\alpha$ with $\tau \sim$0.34 accounts nicely for the 6250\AA~absorption at maximum.
 He I has $\tau \sim$2.5 and  $v_{cont}$= 0 km s$^{-1}$, while SSp at 10days has $v_{cont}$(He I)= 2000 km s$^{-1}$ 
 and  $v_{cont}$(H$\alpha$)= 5000 km s$^{1}$. Na I line is added in the 10days SSp to help the fit of the emission 
 component peaked around 5900\AA. 
\vspace{0.4cm}\\
$*$ {\bf{SN 1984L:}}
\begin{figure}
\includegraphics[height=9cm,width=9cm]{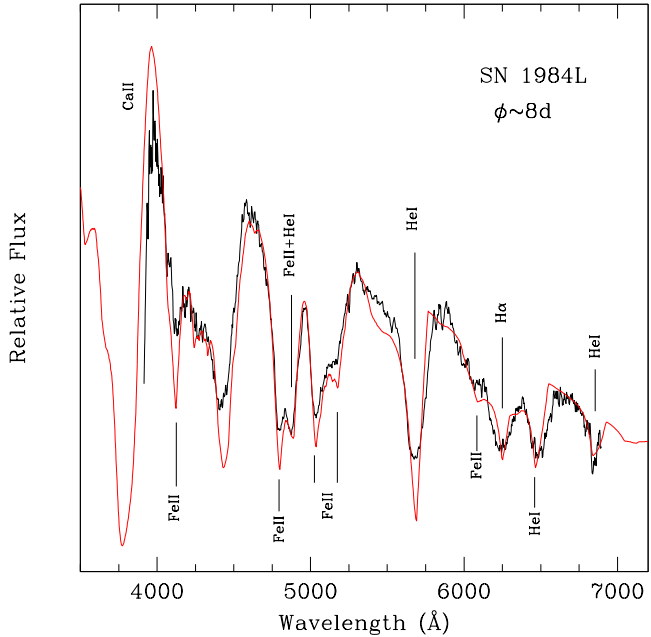}
\caption{SSp fit of SN 1984L compared to the observed spectrum at day 8. Conspicuous line features 
 are shown. }
\end{figure}                                      

Figure 5 compares the observed spectrum at $\sim$8days with an SSp that has $v_{phot} = 8000$ km s$^{-1}$ and 
 T$_{bb}=$8000 K. Ions that are responsible for the most conspicuous absorption features are indicated. Interestingly, 
 He I lines with $v_{cont}$= 2000 km s$^{-1}$ and $\tau =$2.5 , provide a  good match at 5876\AA, 6678\AA~ and 7056\AA. 
 That may indicate that departures from LTE, due to non-thermal effects, are not severe for this Type Ib object. A 
 similar result obtains for the 32day spectrum i.e. ($v_{cont}$(He I)= 2000  km s$^{-1}$ and $\tau =$6.7; Top panel 
 of Fig. 6). The corresponding SSp in Fig. 6 has $v_{phot} = 5000$ km s$^{-1}$ and T$_{bb}=$5600 K.

 The 32days He I profiles show rounded emission components that cannot be matched by a power-law SSp. As we have 
 already mentioned, a detached line in a power-law assumption has a discontinuity in its optical depth (i.e. non-zero 
 optical depth only above the $v_{min}(line)$). When a profile retains a rounded P-Cygni emission component even if 
 its $v_{min}(line)$ is greater than $v_{phot}$, this might indicate two components of optical depth rather than only 
 one above the photosphere. In such cases better fits could be obtained by having gradually decreasing optical depth
  below the detached velocity, instead of a discontinuity. In the present version of SYNOW, only an $e-$folding 
 assumption of the optical depth allows a two-component treatment of a given line. We switch then to the exponential 
 case adopting $v_e$=3000 (see Sect. 2). For He I lines and below $v_{min}$= 7000 km s$^{-1}$ we introduce a second 
 component  with negative $v_e$ ($v_e=-$2000 km s$^{-1}$), such that ``$\tau$'' is continuous at the detachment 
 velocity (i.e. at 7000 km s$^{-1}$). In this case and for similar situations, the line should not be defined as 
 ``detached''. Instead, it has a maximum value of ``$\tau$'' that is not at the photosphere as is normal for an 
 undetached line.
 The bottom panel in Fig. 6 demonstrates that the two-component optical depth reflects more probably the real 
 situation of the He I lines in SN 1984L. The He I line fits seem better relative to the power-law case, although 
 blueward 5500\AA~we obtain an inferior match.
\begin{figure}
\includegraphics[height=9cm,width=9cm]{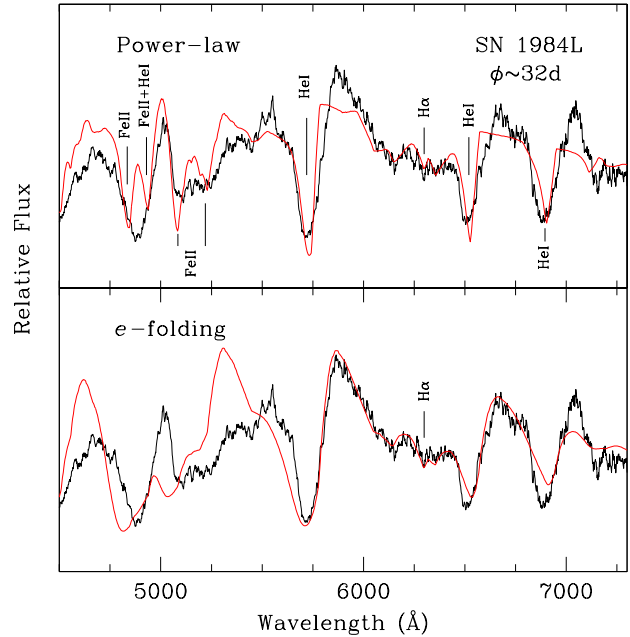}
\caption{SSp fit of SN 1984L compared to the observed spectrum at day 32.
 The lower panel shows the $e-$folding optical depth possibility (see discussion). Conspicuous line 
 features are shown.}
\end{figure}                                      
\begin{figure}
  \includegraphics[height=9cm,width=9cm]{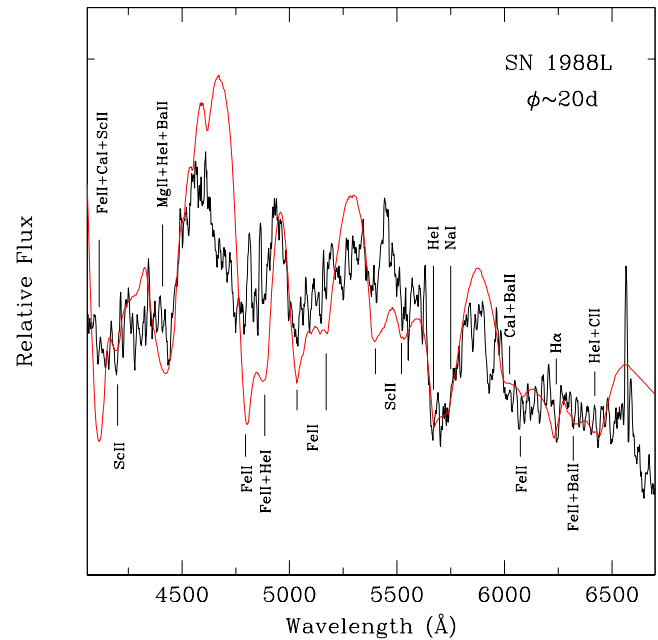}
\caption{ SSp fit of SN 1988L compared to the observed spectrum at day 20.
 Conspicuous line features are shown.}
\end{figure}                                      
 In the 8days SSp, Si II is too blue to account for the 6250\AA. We checked the undetached Ne I 6402\AA~line 
 possibility, assigning it an optical depth of 1.5. 
It gives a profile broader than the observed one. In addition the coverage in wavelength of the spectrum hampers a 
 check  whether Ne I lines produce unwanted features longward of 7000\AA. Ne I lines would need to be non-thermally 
 excited as the case for He I lines in Type Ib (Lucy 1991; Swartz et al. 1993\nocite{Swar93}). With an optical 
 depth of 1.5, Ne I
 would have a departure coefficient from LTE of about 15 (Hatano et al. 1999). Additionally, the low non-thermal 
 effects seen in He I lines of SN 1984L (see above) may argue against Ne I identification.
  C II is also rejected since it will require a $v_{cont}$(CII)= 8000 km s$^{-1}$ while at this phase $v_{cont}$(He I)= 
 2000 km s$^{-1}$ (criterion 4). H$\alpha$ remains then a plausible candidate. In Fig. 5, indeed, 
 $\tau$(H$\alpha$)=0.65 and $v_{cont}$(H$\alpha$)= 7000 km s$^{-1}$  provide a good fit, while at 32days we use 
 $\tau$(H$\alpha$)=0.2 and $v_{cont}$(H$\alpha$)= 7500 km s$^{-1}$  (Fig. 6).  
\vspace{0.4cm}\\
$*$ {\bf{SN 1988L:}}

Figure 7 compares the observed spectrum at $\sim$20d with an SSp that has $v_{phot} = 8000$ km s$^{-1}$ and  
 T$_{bb}=$9000 K. The observed spectrum has been smoothed with a box size of 5. Narrow emission lines due to H II 
 regions are present in the spectrum. The object has been discussed and classified as a Type Ib SN 
 (Filippenko 1988\nocite{Fil88}; Kidger 1988\nocite{Kid88}).
\begin{figure}
  \includegraphics[height=9cm,width=9cm]{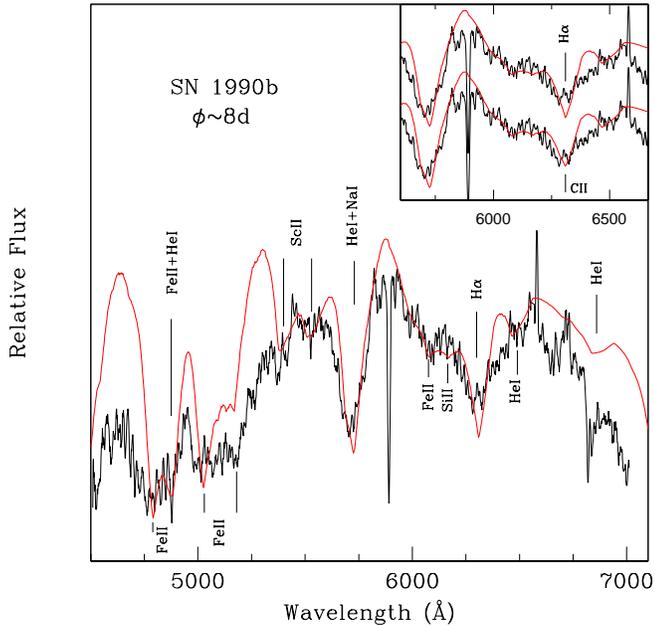}
\caption{SSp fit of SN 1990B compared to the observed spectrum around day 8. 
 Lines that are responsible for the most conspicuous features are also shown. The region around 
 the 6300\AA~trough is zoomed in the window.}
\end{figure}                                      

The strong and broad P-Cygni profile at $\sim$5550\AA~is well matched in our SSp by a blend of Na I and He I lines. 
The He I 5876\AA~accounts for the blue edge of the broad trough, having $v_{cont}$(He I)= 3000 km s$^{-1}$ and 
$\tau$(He I)=1, while undetached Na I has $\tau$(NaI)=3. Ba II lines together with Ca I lines have been introduced 
in the SSp and help the fit blueward $\sim$6000\AA. Undetached Si II 6355\AA~is too blue to account for the trough 
seen at  $\sim$6250\AA. We adopt H$\alpha$ with an optical depth of 0.7 and $v_{cont}$(H$\alpha$)= 8000 km s$^{-1}$. 
 Undetached C II 6580\AA, $\tau$(CII)=0.006, contributes together with He I blueward 6500\AA~(Fig. 7). SN 1988L may 
be regarded as an intermediate Ib/c object rather than a typical Type Ib.  
\vspace{0.4cm}\\
$*$ {\bf{SN 1990B:}}

An extensive study of SN 1990B observations has been presented by Clocchiatti et al. (2001)\nocite{Cloc01}. 
 The authors pointed out 
the red character of the object. This fact is supported in the present analysis, see below, by means of the estimated 
continuum temperatures that seem much lower than values at similar phases for the other Type Ib-c sample objects.
  
The spectrum at day 8 with the best fit SSp is presented in Figure 8. The SSp has $v_{phot} = 8500$ km s$^{-1}$ and  
T$_{bb}=$5400 K. Ions that are responsible for the most conspicuous features are indicated. The fit parameters have
 been modified many times in order to investigate line identification possibilities. The strong P-Cygni profile
 around 5700\AA~is a blend of undetached Na I and detached He I 5876\AA~($v_{cont}$(He I)= 1500 km s$^{-1}$ and 
 $\tau$(He I)=0.55). The expelled helium, with $v_{cont}$(He I)= 1500 km s$^{-1}$, accounts nicely for the weak 
 absorption near 6500\AA. Undetached Si II 6355\AA~is considered to fit the weak absorption at $\sim$6200\AA.
 The H$\alpha$ line, $v_{cont}$ (H$\alpha$)= 4500 km s$^{-1}$ and  $\tau$(H$\alpha$)=0.5, is introduced to account 
for the broad observed absorption trough around 6300\AA. The window in Figure 8 displays a zoom of the 6300\AA ~region,
 together with the C II fit possibility ($\tau$(CII)=0.0025). The fit looks as good as H$\alpha$, however it would 
need to  be expelled with 3000 km s$^{-1}$ 
more than the He I lines (i.e. $v_{cont}$(CII)= 4500 km s$^{-1}$), and hence is ruled out (criterion 4). 
\begin{figure}
  \includegraphics[height=9cm,width=9cm]{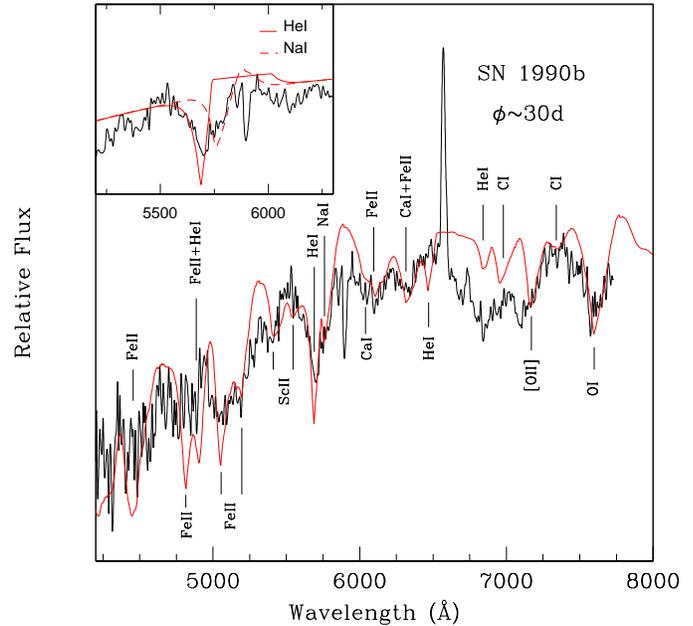}
\caption{SSp fit of SN 1990B compared to the observed spectrum around day 30. 
 Lines of conspicuous features are shown. The region around the 5700\AA
 ~trough is zoomed in the window. Na I D and He I 5876\AA~individual contributions are also displayed.}
\end{figure}                                  
\begin{figure}
\includegraphics[height=9cm,width=9cm]{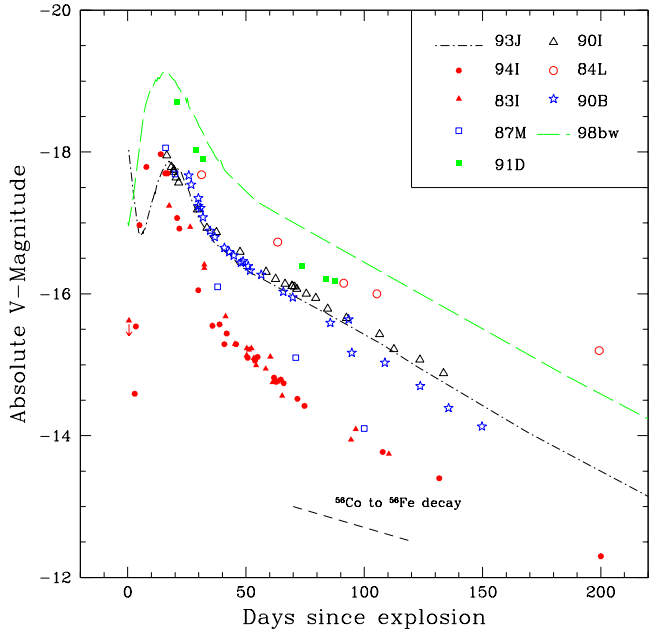}
\caption{ The V-band light curve of SN 1990B compared to a sample of Ib-c ``V-light curves'' (See 
text for details).}
\end{figure}                         

The second spectrum, around day 30, is compared to an SSp that has $v_{phot} = 7000$ km s$^{-1}$ and  T$_{bb}=$4000 K 
(Fig. 9). The presence of He I 5876\AA~is supported by two facts, first by the wide nature of the absorption trough 
that cannot be caused only by the undetached Na I D. Second, by the emergence of a ``$bump$'' around 5760\AA, in the 
transition slope redward of the deep absorption trough at $\sim$5700\AA. A closer view of the ``He I+Na I'' region is 
provided in the window of Fig. 9. The window demonstrates the contribution of both Na I D and He I 5876\AA~separately. 
The match is good and reproduces nicely the total feature seen in the observed profile. This tends therefore to confirm
 the presence of helium in the ejecta of this object, even though it does not exhibit a full and clear set of helium
 absorption lines. He I lines have $v_{cont}$(He I)= 3000 km s$^{-1}$ and $\tau$(He I)=0.8. On the other hand, Fe II 
lines with $\tau$(FeII)=12, provide good fit in the blue part of the spectrum. In addition Fe II contributes 
 significantly in the absorption at $\sim$6320\AA. The presence of H$\alpha$ is therefore not needed since the
 Fe II contribution is sufficient. Nevertheless if H$\alpha$ were present it would have a $v_{cont}$(H$\alpha$)= 
 4000 km s$^{-1}$.

 Based on its spectroscopic properties, especially the weak He I lines, SN 1990B was re-classified as a Type Ic object 
(Clocchiatti et al. 2001). It was first classified as Type Ib SN (Kirshner et al. 1990)\nocite{Kirsh90}. 
 Clocchiatti et al. (2001) 
 presented an extensive set of well-sampled photometry. In Figure 10, we show the absolute V$-$light curve of SN 1990B
 compared to other SNe Ib-c. The  comparison
 highlights the similarity with Type IIb 1993J, SN Ib 1990I and SN Ib 1991D. The maximum brightness is comparable to 
 those of SNe 1990 I, 1987M and 1994I and intermediate between the bright SNe 1999bw and 1991D and the faint SN 1983I. 
 Moreover the figure reveals a similar decline rate, from maximum to reach the exponential decay, of SNe 1990B, 1990I, 
 1993J, 1991D and 1998bw and indicates 
similar peak-to-tail contrast, whereas SNe 1983I, 1987M and 1994I display narrower peak widths and greater peak-to-tail 
 contrast. The events with steeper decline rates (i.e. narrow widths) and greater peak-to-tail contrast are classified 
 as Type Ic SNe, while SNe 1990I and 1991D are Type Ib events. This fact may point to a photometric behaviour closer 
 to Type Ib rather than Type Ic.
 We regard, therefore, SN 1990B as an intermediate Type Ib/c event.
\vspace{0.4cm}\\
$*$ {\bf{SN 1991ar:}}
\begin{figure}
  \includegraphics[height=10cm,width=9cm]{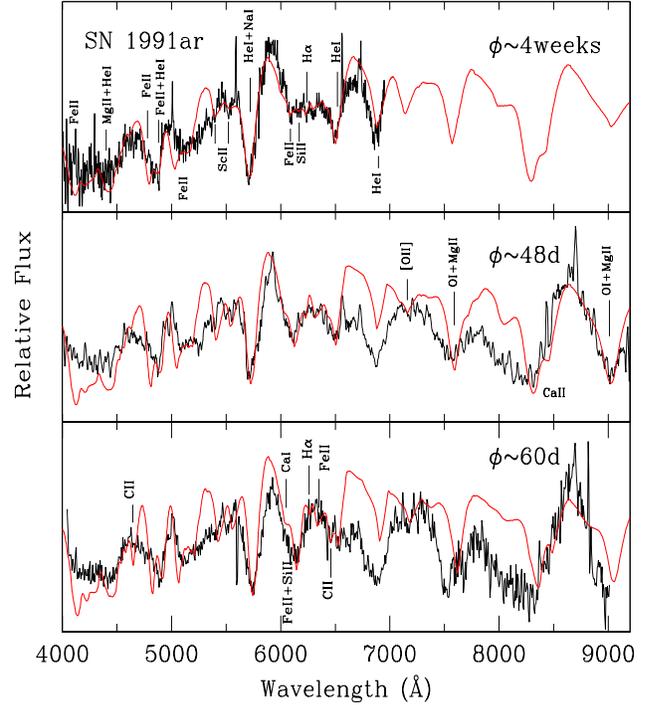}
\caption{SSp fit of SN 1991ar compared to the observed spectra around 4 weeks (top panel), 48days 
(middle panel)
 and at 60days (lower panel). Lines that are responsible for the most conspicuous features are 
reported.}
\end{figure}                                      

Figure 11 displays the available spectra of SN 1991ar at 3 different phases, namely 28days, 48days and 60days. The 
 first spectrum is compared to an SSp that has $v_{phot} = 8000$ km s$^{-1}$ and T$_{bb}=$4400 K. The broad  P-Cygni 
profile with the minimum trough centered at $\sim$5720\AA~is well fitted by the ``He I+Na I D'' blend. Although the 
spectrum covers only the $4000-7000$\AA~wavelength range,
 it shows distinctly the He I series, indicating a  Type Ib nature. The He I lines are undetached and have 
 $\tau$(He I)=10. One particularity of this spectrum is the flat behaviour of the $\sim$6070$-$6450\AA~range. We checked
 different ion combinations, and the best fit to that zone is reached including lines of Fe II, Si II and H$\alpha$. 
 This latter has  $v_{cont}$(H$\alpha$)= 8000 km s$^{-1}$ and $\tau$(H$\alpha$)=0.32.

 For the 48day spectrum, the match is good with the SSp having  $v_{phot} = 7000$ km s$^{-1}$ and T$_{bb}=$4800 K 
 (Fig.11;  middle panel). The He I lines are now slightly detached with $v_{cont}$(He I)= 1000 km s$^{-1}$ and 
 $\tau$(He I)=4.8 Lines of O I and Mg II both contribute to the observed features around 7600\AA~and 9030\AA~
($\tau$(OI)=3.5 and $\tau$(MgII)=5), while the Ca II IR triplet is produced by $\tau$(CaII)=500. The weak feature
 labeled as being due to H$\alpha$ corresponds to $v_{cont}$(H$\alpha$)= 9000 km s$^{-1}$ and $\tau$(H$\alpha$)=0.4.

 The bottom panel in Figure 11 shows the 60day spectrum compared to an SSp that has $v_{phot} = 6000$ km s$^{-1}$ and
  T$_{bb}=$4600 K. The $\sim$6080$-$6500\AA~range loses the flat behaviour seen in the first spectra and a round 
 emission profiles start to form. The match in that part of the spectrum with the SSp is acceptable. The profile is 
 mainly due to lines of Fe II, Si II, C II, H$\alpha$ and Ca I. The He I fit corresponds to
 $v_{cont}$(He I)= 1000 km s$^{-1}$ and $\tau$(He I)=4.4, while H$\alpha$ has $v_{cont}$(H$\alpha$)= 8000 km s$^{-1}$ 
 and $\tau$(H$\alpha$)=0.28. Note that the two weak troughs due to the Sc II lines at 5527 and 5661\AA~appear near
 5500\AA~in all 3 spectra. Another notable feature is the deficiency, in flux, of the observed spectra, in the middle 
 and bottom figures, relative to the synthetic ones especially redward of $\sim$6500\AA. This is possibly
 because the supernova already enters a transition phase to the nebular epoch.     

 Surely the evidence of H$\alpha$ should not be taken as definite, although the corresponding weak absorptions in the 
 SSp nicely matches the observed features. 
\vspace{0.4cm}\\
$*$ {\bf{SN 1991D:}}
\begin{figure}
  \includegraphics[height=9cm,width=9cm]{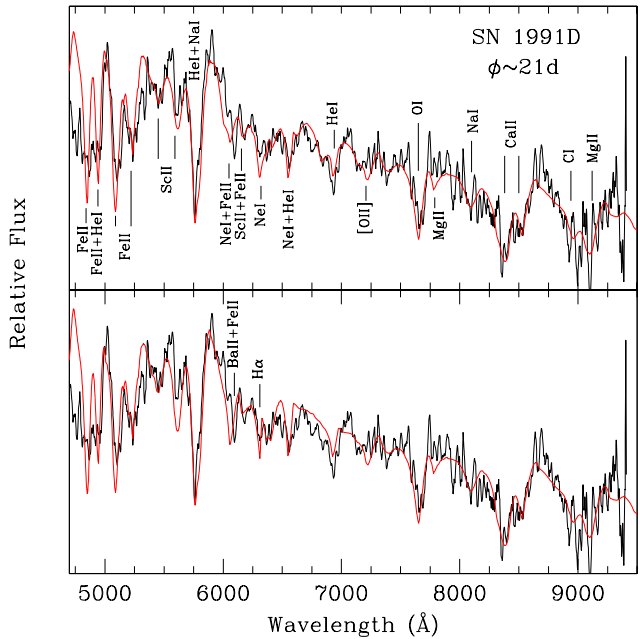}
\caption{ SSp fit of SN 1991D compared to the observed spectra around 3 weeks. Lines that are responsible for the most 
 conspicuous features are reported. The top panel illustrates the Ne I possibility, while H$\alpha$ case is displayed 
 in the lower panel.}
\end{figure}                                      

SN Ib 1991D data has been presented and studied by Benetti et al. (2002).
The object, with its narrow features, seems having lower velocities compared to other Type Ib events at similar phases.
 When analyzing the $\sim$10day spectrum by means of SYNOW code, the authors obtained an improved match introducing 
undetached Ne I lines with respect to the H$\alpha$ SSp fit. Figure 12 illustrates this possibility for the 21days
 spectrum. The displayed SSp has $v_{phot} = 4600$ km s$^{-1}$ and  T$_{bb}=$7000 K. The He I lines are evident 
through our SSp ($v_{cont}$(He I)= 1400 km s$^{-1}$ and $\tau$(He I)=1.8). In addition, the Na I D feature 
($\tau$(NaI)=4) contributes to the He I 5876\AA~broad P-Cygni profile. The presence of Na I lines is consistent 
 with the good fit around 8100\AA. The most conspicuous line absorption features are indicated in the figure. In the 
 top panel of Figure 12, undetached  Ne I lines are included in the SSp ($\tau$(NeI)=2). The match in the whole 
 spectrum is quit good. In the bottom panel the H$\alpha$ possibility is tested. Except for adding Ba II lines in 
 order to help the fit at $\sim$6100\AA, the other ion parameters are kept unchanged. The fit to the observed 
 absorption near 6300\AA~ with H$\alpha$, $v_{cont}$(H$\alpha$)= 7400 km s$^{-1}$ and $\tau$(H$\alpha$)=0.46, is 
 slightly better compared to the Ne I case 
(Top panel). However the SSp, in the bottom panel, does not account for the observed features near 6630\AA~and 
 6840\AA~as does NeI lines in the SSp of the top panel. Ne I remains hence a strong candidate in this Type Ib
 object.

 It is worth noting here that at a similar phase, i.e. $\sim$21days, SN Ib 1990I had a photospheric velocity of 
 9500 km s$^{-1}$.          
\vspace{0.4cm}\\
$*$ {\bf{SN 1991L:}}

We study a spectrum dated about one month after maximum (Fig. 13). The spectrum has been smoothed with a box size of 3. 
The classification of this event based only on the shown spectrum is very tentative. Indeed the spectrum shows a 
 short wavelength part reminiscent of Type Ia spectra. However, nebular spectra of SN 1991L have been found to display 
 emission features that are normally found in Type Ib/c events (Gomez $\&$ Lopez, 2002)\nocite{Gom02}. 
 The illustrated best fit 
 SSp has $v_{phot} = 5000$ km s$^{-1}$ and  T$_{bb}=$6000 K. The He I lines are not so obvious, although an undetached 
 He I 5876\AA~line, $\tau$(He I)=2, provides a good fit to the observed absorption trough at $\sim$5880\AA. Na I D 
 could contribute to that feature, although it would provide an absorption component slightly redshifted. We regard 
 SN 1991L as a transition Type Ib/c object.

Similarly to SN 1991D, we tested the H$\alpha$ and Ne I possibilities as a candidates for the feature near 6290\AA. 
 Figure 13 shows both cases. The included Ne I lines in the upper panel have $\tau$(He I)=1, while in the lower panel 
 H$\alpha$ has $v_{cont}$(H$\alpha$)= 8000 km s$^{-1}$ and 
$\tau$(H$\alpha$)=0.34. Some features due to Ne I lines appear similar to the observed ones, which are not accounted 
 for the H$\alpha$ case (lower panel; Fig. 13). Compared to the lower panel, Ne I lines also improve the fit of the 
 broad emission component of the He I 5876\AA~P-Cygni profile. Ne I remains therefore an alternative possibility 
 to the H$\alpha$ identification.
\begin{figure}
  \includegraphics[height=9cm,width=9cm]{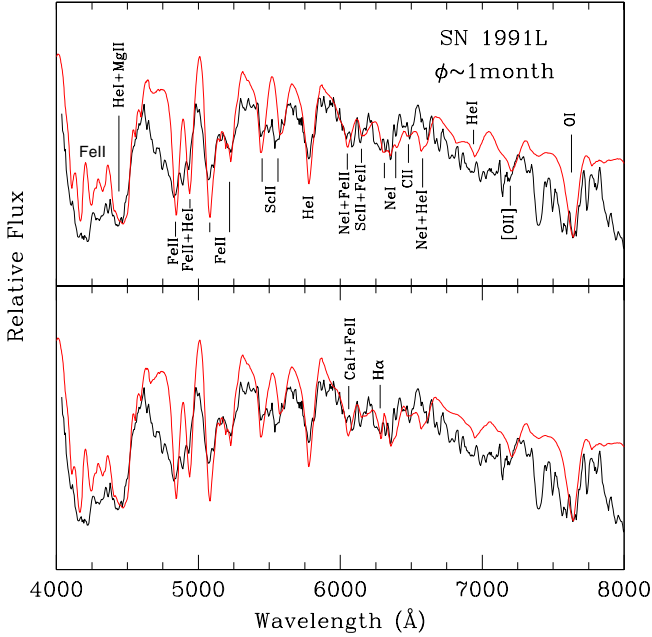}
\caption{ SSp fit of SN 1991L compared to the observed spectra around 1 month. Lines that are responsible for the 
 most conspicuous features are reported. Top panel illustrates the Ne I  possibility, while H$\alpha$ case is 
 displayed in the lower panel.}
\end{figure}                                      
\vspace{0.4cm}\\
$*$ {\bf{SN 1997dc:}}

The best SSp fit, $v_{phot} = 5000$ km s$^{-1}$ and  T$_{bb}=$4200 K, is displayed in Figure 14 together with the 
 observed spectrum ($\sim$4 weeks since maximum). On the one hand, the match with He I series, $v_{cont}$(He I)= 
 2000 km s$^{-1}$ and $\tau$(He I)=2, is evident, confirming the classification as a Type Ib event. On the other hand 
 and in order to obtain an improved fit to the He I 5876\AA~P-Cygni profile, undetached Na I D with $\tau$=5 is needed. 
 The identification of Na I is consistent with the absorption profile seen near 8090\AA. Ca II IR triplet is unusually 
 strong and the match to the synthetic one is rather poor. Sc II with $\tau$=2 is responsible for absorption features
 blueward of the He I 5876\AA, and contributes as well, with Fe II, near 6160\AA.

A zoomed view of the 5600$-$6600\AA~region is shown in the window of Figure 14. The adopted SSp fits well the observed 
features. Various ion combinations have been tested, especially to explain the weak absorption at $\sim$6290\AA. The 
 best fit that would not introduce unwanted features in the rest of the spectrum and as well be in agreement with our 
 previous criteria (Sect. 3.1) is attributed to H$\alpha$ having  $v_{cont}$(H$\alpha$)= 8000 km s$^{-1}$ and 
$\tau$(H$\alpha$)=0.2.
\begin{figure}
  \includegraphics[height=9cm,width=9cm]{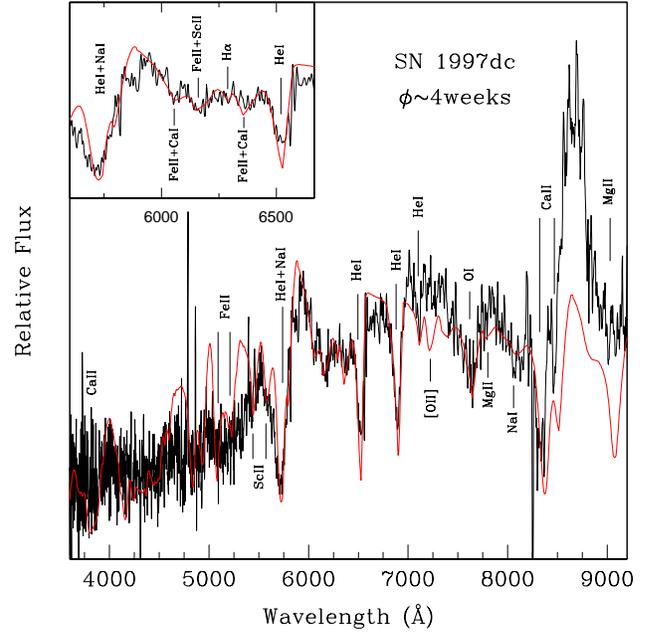}
\caption{ SSp fit of SN 1997dc compared to the observed spectrum at about 4 weeks. Lines of conspicuous 
 features are shown. The region around the 6300\AA~ weak trough is zoomed in the window (see text).}
\end{figure}                                      
\vspace{0.4cm}\\
$*$ {\bf{SN 1998dt:}}

Spectra at 2 phases, $\sim$8days and $\sim$33days, are shown in Figure 15 and compared with the best fitting synthetic 
 spectra. The SSp in the upper panel corresponds to  $v_{phot} = 9000$ km s$^{-1}$ and  T$_{bb}=$5600 K, while for 
 $\sim$33days, the SSp has  $v_{phot} = 9000$ km s$^{-1}$ and  T$_{bb}=$5000 K (lower panel). The SSp nicely match 
 the  features in the observed spectra. The narrow emission near 6565\AA~is due to H$\alpha$ from H II region.

The reference Ca II line (i.e. at 3933\AA) in the upper panel has an optical depth of 500. The corresponding 
 Ca II H$\&$K and IR Ca II triplet (8542, 8662, 8498\AA) SSp both fit well the observed broad P-Cygni profiles. For 
 the 33day spectrum however, the Ca II infrared triplet ($\tau$=500) presents a deficit with respect to the observed 
 profile even though for the absorption
 part the fit is still acceptable. The O I 7773\AA~with $\tau$=0.5 accounts for most of the observed feature at both 
 phases (top and lower panels).  
 
 The He I lines are clearly evident for both phases, with the main difference that while at $\sim$8days He I lines 
 are detached ($v_{cont}$(He I)= 2000 km s$^{-1}$ and $\tau$(He I)=4), they have non-zero optical depths starting at 
 the photosphere for the $\sim$33day spectrum ($v_{cont}$(He I)= 0 km s$^{-1}$ 
and $\tau$(He I)=10). Note here that for both phases He I 5876\AA~is sufficient to fit the observed P-Cygni profile. 
 There is no need to include Na ID line. The slope redward of the He I 5876\AA~emission component is nicely accounted 
 for by lines of Fe II, Ca I and H$\alpha$ at  $\sim$8days, while for the 33day spectrum 
 we introduce, in addition, lines of Sc II ($\tau$=2). The presence of Sc II is supported by the double absorption 
 features near 5390 and 5500\AA.

The best SSp fit of the observed 8day spectrum includes only 6 elements, namely Fe II, He I, Ca I, Ca II, O I and H I. 
 H$\alpha$, with $v_{cont}$= 8000 km s$^{-1}$ and $\tau$=0.3, accounts for the weak absorption near 6200\AA. A similar 
 but weaker absorption is seen in the 33day spectrum as well. We fit it with H$\alpha$ that has  $v_{cont}$= 8000 
 km s$^{-1}$ and $\tau$=0.2. It is clear that even though the fit is convincing, the presence of H$\alpha$ is not 
 definite. This limitation occurs whenever we deal with absorption features that are not so deep with respect to the 
 continuum. 
\begin{figure}
 \includegraphics[height=9cm,width=9cm]{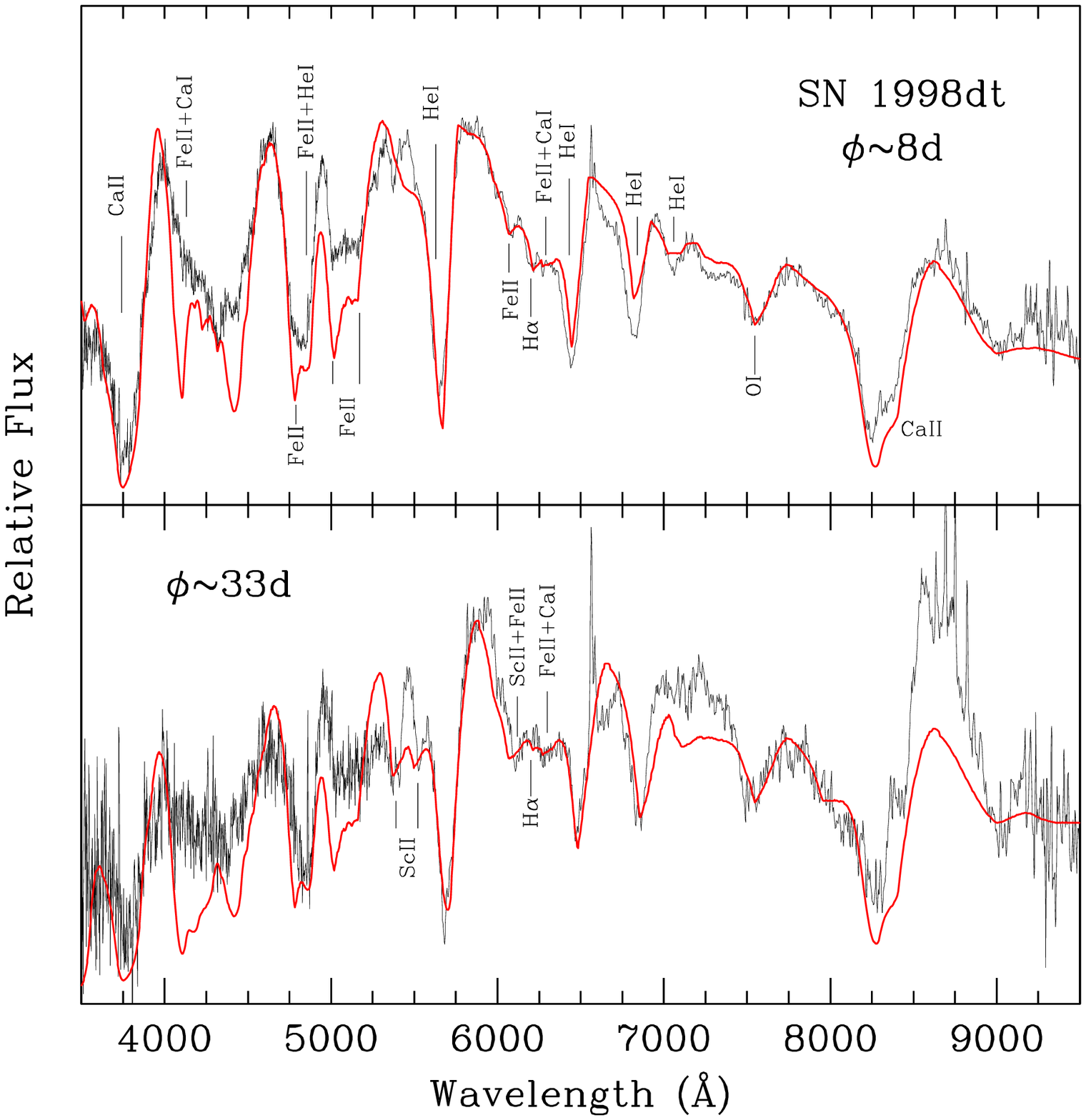}
\caption{SSp fit of SN 1998dt compared to the observed spectra at 8days (upper panel) and at 33days (lower panel). 
 Lines of conspicuous features are indicated.}
\end{figure}      
\vspace{0.4cm}\\
$*$ {\bf{SN 1998T:}}

Spectra of SN 1998T at two different phases, $\sim$ 21days and 42days, are displayed in Figure 16 and compared to 
 synthetic spectra having $v_{phot} = 6000$ km s$^{-1}$ and  T$_{bb}=$5400 K (upper panel), and $v_{phot} = 5800$ 
km s$^{-1}$ and  T$_{bb}=$5200 K (lower panel). The spectra are highly contaminated by weak emission features from 
 an H II region. The absorption troughs caused by He I series are prominent, indicating a familiar Type Ib appearance. 
The He I line best fit at 21days corresponds to $v_{cont}$(He I)= 3000 km s$^{-1}$ and $\tau$(He I)=1.15, while at 
 42days $v_{cont}$(He I)= 2200 km s$^{-1}$ and $\tau$(He I)=0.75. The optical depth of the undetached Ca II is 
 120 (upper panel) and 140 (lower panel). The corresponding features, i.e of Ca II, fit well the observed ones 
 although an excess in  the emission P-Cygni components is visible, as is the case in most advanced photospheric 
 spectra. C I, with $\tau$=0.3, helps fit the absorption feature near 8930\AA. Fe II lines are unusually weak for 
 SN 1998T. Indeed, optical depths of 1 and 5 are respectively adopted for our best synthetic fits at $\sim$21days
  and $\sim$42days.
\begin{figure}
 \includegraphics[height=9cm,width=9cm]{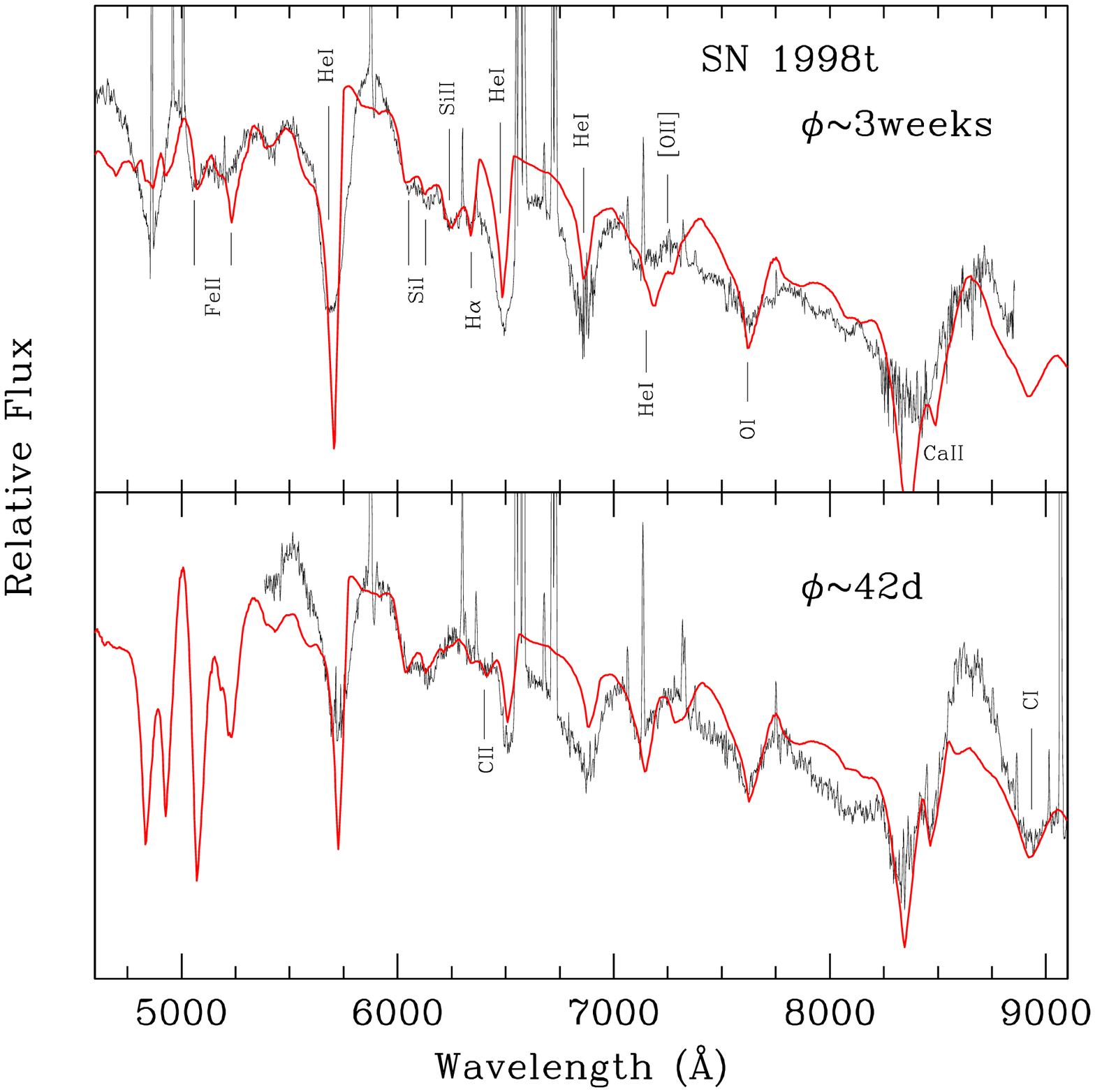}
\caption{SSp fit of SN 1998T compared to the observed spectra at 3 weeks (upper panel) and at 42days  
(lower panel). Lines of conspicuous features are indicated.}
\end{figure}      

 We checked different combinations for the 6000$-$6500\AA~wavelength range, testing the possible candidates and 
 keeping in mind our departure criteria (Sect. 3.1). For the 21day spectrum, the best fit includes Si I($\tau$=0.03), 
 Si II($\tau$=0.6), Fe II and H$\alpha$. This latter has $v_{cont}$ (H$\alpha$)= 4500 km s$^{-1}$ and
  $\tau$(H$\alpha$)=0.34. When testing Ne I lines one obtains an improved fit blueward of the He I 7065\AA~feature, 
 but then unwanted profiles in the 6000$-$6500\AA~region would be introduced. At low temperatures, an Si I 
 identification is plausible in Type Ib objects within LTE assumption with low optical depths (Hatano et al. 1999). 
 Si II however, is expected (in LTE) to have an  optical depth of about 10 at temperatures similar to that for the 
 21day spectrum (helium-rich composition; Hatano et al. 1999). This would mean that we might have mis-identified the
 absorption feature near 6250\AA, or we have a departure from LTE in the Si II line. Note that for Type II SNe, and at 
 similar temperatures, Si II has an optical depth as low as 0.1 (hydrogen-rich composition; Hatano et al. 1999). The 
 remaining possibility is that the 6250\AA~is due to H$\alpha$ rather than Si II. In this case hydrogen would have 
 $v_{cont}$(H$\alpha$)= 9000 km s$^{-1}$, but then the absorption near 6340\AA~remains unaccounted for. To fit this 
 latter with C II, we need then to impose a velocity of about $v_{cont}$=6000 km s$^{-1}$, which means that carbon is 
 expelled 3000 km s$^{-1}$ greater than helium, contradicting our criteria (Sect. 3.1).  The most acceptable 
 combination is then the one illustrated in the top panel of Figure 16.

The situation for the 42day spectrum is more difficult as the spectrum is noisier. Our best fit includes 
 Si I($\tau$=0.025), Si II($\tau$=0.4), Fe II and C II ($\tau$=2.3$\times$10$^{-4}$). We prefer C II, having in 
 this case similar $v_{cont}$ as He I, rather than adopting H$\alpha$. First, because hydrogen would need then to 
 be expelled at velocities lower than helium, and second because we remain consistent with our criteria. 
We propose that it was H$\alpha$ at 21days which was then overwhelmed by C II later as the photosphere recedes 
 because of the envelope expansion.    
\vspace{0.4cm}\\
$*$ {\bf{SN 1999di:}}
\begin{figure}
\includegraphics[height=9cm,width=9cm]{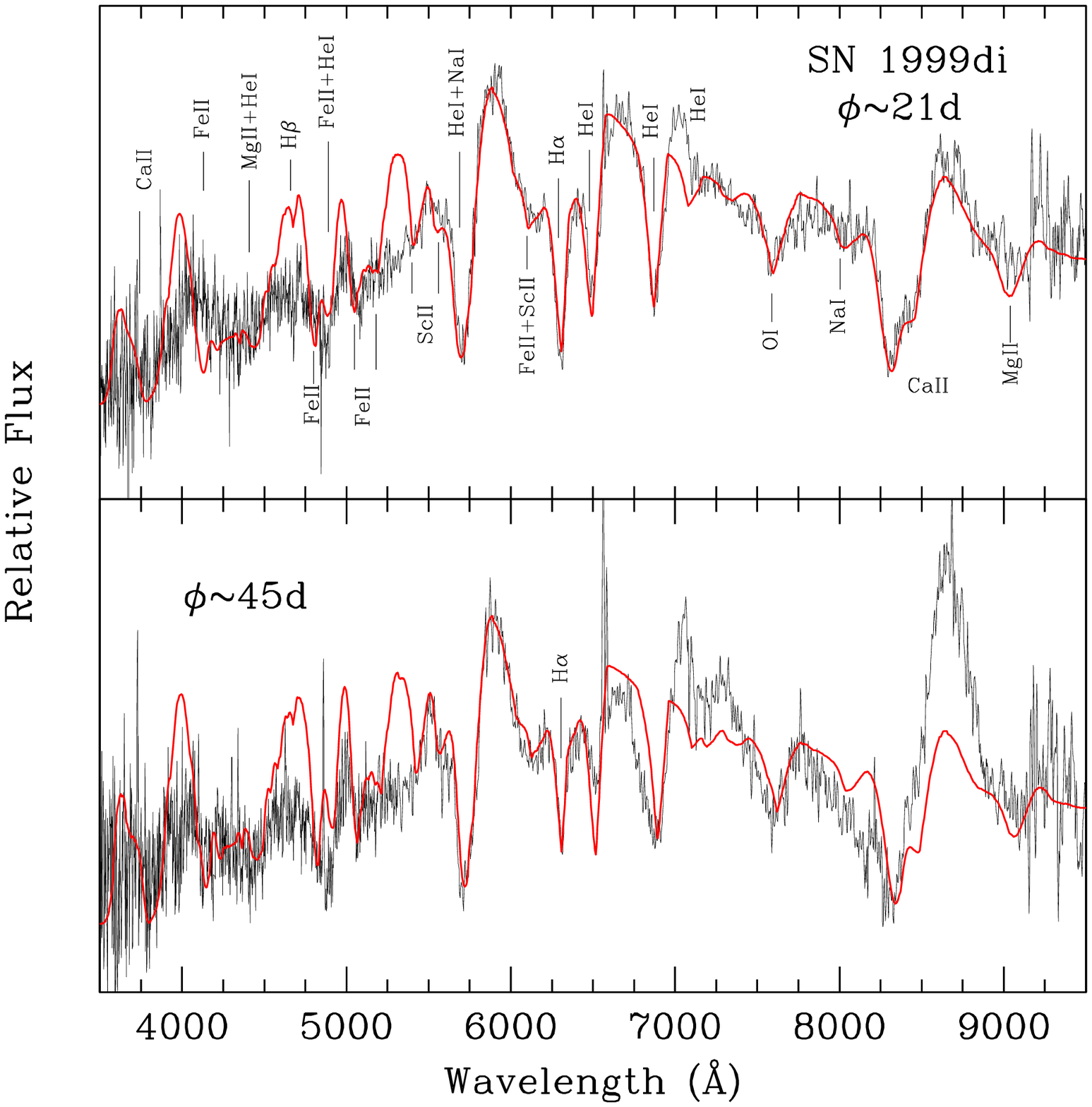}
\caption{SSp fit of SN 1999di compared to the observed spectra at 21days (upper panel) and at 
 45days (lower panel). Lines of conspicuous features are indicated.}
\end{figure}                   

In Figure 17, the observed spectra of SN 1999di at $\sim$21days and 45days are compared to the computed synthetic 
 spectra. The spectra are synthesized adopting  $v_{phot} = 7000$ km s$^{-1}$ and  T$_{bb}=$4800 K (upper panel) 
 and $v_{phot} = 6000$ km s$^{-1}$ and  T$_{bb}=$5200 K (lower panel). The match is good in the overall spectral 
 shape. The fit to the He I lines is obvious. The parameters used are $v_{cont}$(He I)= 1500 km s$^{-1}$ and 
 $\tau$(He I)=12.6 for the 21day spectrum, while at 45days the He I optical depth diminishes to $\tau$(He I)=10, 
 keeping the same contrast velocity (i.e. 1500 km s$^{-1}$).
 For the 21day spectrum the optical He I lines (i.e. at 5876, 6678 and 7065\AA) are simultaneously well accounted 
 for by their corresponding absorption troughs in the SSp. Even the He I 7281\AA~line accounts for an observed 
 absorption. At day 45, however, the SN displays a shallower He I 6678\AA~compared to the He I lines at 5876\AA~and 
 7065\AA. At both phases, Na ID 5893\AA~is blended with He I 5876\AA. The absorption near 8030\AA~is also accounted 
 for by lines of Na I. The features blueward of the broad ``He I+Na ID'' P-Cygni profile are attributed to Sc II 
 lines ($\tau$=2 for both epochs). Sc II also helps the fit redward of the emission component of the blended 
 ``He I+Na ID'' P-Cygni profile.
\begin{figure}
  \includegraphics[height=9cm,width=9cm]{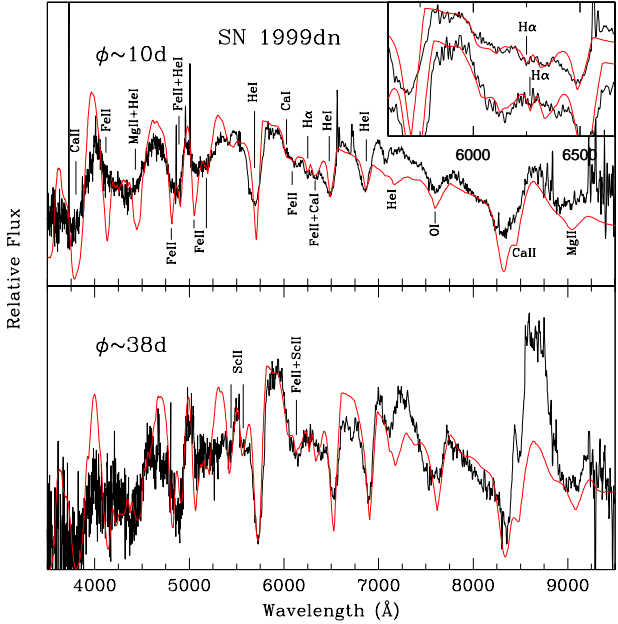}
\caption{SSp fit of SN 1999dn compared to the observed spectra at 10days (upper panel) and at 38days (lower panel). 
 Lines of conspicuous features are indicated. The region around the 6300\AA~weak trough is zoomed in the window (see text).}
\end{figure}                   

The O I 7773\AA~line, with $\tau$=1, accounts for most of the observed feature. Note however that the line of 
 Mg II at 7890\AA~contributes in some cases to the red edge of the O I 7773\AA~line. In the present case this 
 contribution is not so relevant. In the upper panel, the Ca II IR triplet, with an optical depth of 700, fits very
 well the observed profile. At day 45, the fit is under-estimated especially for the emission part, which may indicate
  a transition to the nebular phase. The reproduced Ca II H$\&$K line is also strong in both synthetic spectra.  
 The issue of the 6300\AA~feature is particular in this object, because it is very deep and different from features 
 seen in ``normal'' Type Ib SNe. We have tested various identifications for the deep feature. The best 
 fit that would be consistent with our criteria attributes the trough to H$\alpha$ with  $v_{cont}$(H$\alpha$)= 
 5000 km s$^{-1}$ and $\tau$(H$\alpha$)=3 (upper panel), and $v_{cont}$(H$\alpha$)= 6000 km s$^{-1}$ and 
 $\tau$(H$\alpha$)=1.6. With these parameters and in addition to the almost perfect H$\alpha$ fit, the SSp
  accounts as well for the notch\footnote{The word "notch" referes to a narrow and weak absorbtion feature.}
 seen near 4665\AA, assigning it to H$\beta$. The H$\alpha$ trough is reminiscent 
 of what is seen in Type II and IIb objects, with the difference that in these latter types H$\alpha$ displays a 
 complete P-Cygni profile (i.e. with both conspicuous emission and absorption components). This is understandable 
 within the context of ``detachment'' concept. Indeed, Branch et al.(2002) have shown that an H$\alpha$ P-Cygni 
 profile loses its obvious emission component when it is highly
 detached. If correct, we would then expect an 
 increasing value of $v_{cont}$(H$\alpha$) as we go from Type II to IIb to Ib SNe. This point will be emphasized 
 and discussed at the end of the paper.    
\vspace{0.4cm}\\
$*$ {\bf{SN 1999dn:}}

Two spectra of SN 1999dn, observed at 10 and 38 days after maximum light, are compared in Figure 18 to synthetic 
 spectra that have $v_{phot} = 7000$ km s$^{-1}$ and  T$_{bb}=$5800 K (upper panel) and $v_{phot} = 6000$ 
 km s$^{-1}$ and  T$_{bb}=$5400 K (lower panel). As shown in both panels, the prominent lines of He I 5876, 6678 
 and 7065\AA~are clearly accounted for, indicating a typical Type Ib nature. The fit to the He I reference line 
 corresponds to $v_{cont}$(He I)= 2000 km s$^{-1}$ and $\tau$(He I)=1.9 at day 10, and $v_{cont}$(He I)= 
 1000 km s$^{-1}$ and $\tau$(He I)=14.5 at day 38.
 A spectral analysis of SN 1999dn has been also presented by Deng et al. (2000). The authors discussed line 
 identifications for 3 photospheric spectra (at -10, 0 and 14 days from maximum). The observed trough at 
 $\sim$6250\AA~was first blended with He I 6678\AA~for the -10day spectrum, while around maximum light the two troughs 
 become distinctly isolated, giving rise to a double absorption profile (Fig. 2; Deng et al. 2000), very similar 
 to what is seen in SN 1990I (Fig. 1), although the two events evolve afterwards in different ways, with the 
 $\sim$6250\AA~feature being less deep in SN 1999dn than in SN 1990I later on. The authors found it difficult to 
 attribute the minimum near 6250\AA~to Si II 6355\AA. They argued that it was H$\alpha$ first, before maximum, that 
 becomes blended and overwhelmed by C II line in later spectra. For the
 two early spectra, i.e. at -10days and 
 maximum, C II 6580\AA~ provides a fit as good as H$\alpha$, however C II 6580\AA~is assigned a minimum velocity much 
 higher than the one attributed to He I. Highly detached C II lines are surely improbable in this class of event. Even 
 in their late spectrum (i.e around 14days), the C II has a value of $v_{cont}$ higher than that of He I.

 We analyzed various possibilities among line candidates for our 10 and 38 day spectra. The wavelength range of 
 interest, $\sim5800-6600$\AA, is zoomed in the window of Figure 18. We find that the best fits which would 
 reproduce the weak absorption features and not contradict the fit criteria, 
 include lines of He I, Fe II, Ca I and H$\alpha$. This latter accounts nicely for the absorption near 6260\AA~with 
 $v_{cont}$(H$\alpha$)= 8000 km s$^{-1}$ and $\tau$(H$\alpha$)=0.24 (at 10days), and 
 $v_{cont}$(H$\alpha$)= 8000 km s$^{-1}$ and $\tau$(H$\alpha$)=0.34 (at 38days).
\vspace{0.4cm}\\
$*$ {\bf{SN 1999ex:}}

\begin{figure}
  \includegraphics[height=9cm,width=9cm]{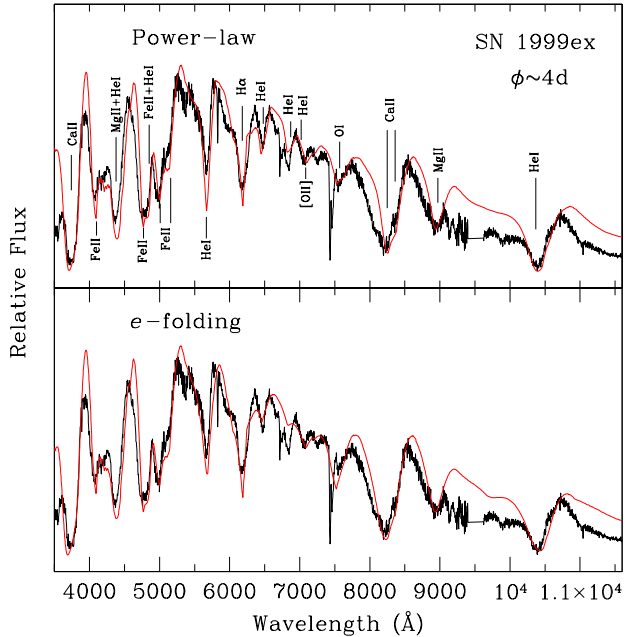}
\caption{SSp fit of SN 1999ex compared to the observed spectrum at 4days. The observed IR spectrum is 
 combined with the optical one in order to allow the He I 10830\AA~identification (see discussion). The 
 Lines of conspicuous features  are reported.}
\end{figure}
 A good set of early data for SN 1999ex, spectra and photometry, have been presented by 
 Hamuy et al. (2002)\nocite{Ham02}. The 
 quality of the photometry, starting well before maximum light, offers a unique possibility of looking at the early 
 behaviour of light curves. Indeed, the early $``dip''$ seen in the $''U''$ and $''B''$ light curves is interpreted 
 as being due to the shock breakout, supporting the present belief that Type Ib-c SNe are the outcome of core 
 collapse in massive stars rather than thermonuclear disruption of white dwarfs (Hamuy et al. 2002; Stritzinger 
 et al. 2002\nocite{Stri02}). Because of weak optical He I lines, the object was classified as an 
 intermediate case between Ib and Ic
  SNe. The evident trough at $\sim$6250\AA~was attributed to Si II 6355\AA~(Fig. 4; Hamuy et al. 2002). The authors 
 also presented three infra-red spectra. 

 We analyze, by means of synthetic spectra, line identifications in this interesting object. The two observed 
 spectra, around 4 and 13 days, are compared with our best fit SSp that have  
 $v_{phot} = 10000$ km s$^{-1}$ and  T$_{bb}=$5800 K (Fig. 19) and  $v_{phot} = 7000$ km s$^{-1}$ and
  T$_{bb}=$5600 K (Fig. 22).  In Figure 19, we combined the IR spectrum with the optical one in order to check the 
 consistency of He I identification. In fact, at 4days, assigning the following parameters: 
 $v_{cont}$(He I)= 1000 km s$^{-1}$ and $\tau$(He I)=2.35, we obtain a good match to 
the observed He I profiles. On the one hand, a strong support for the presence of He I lines comes from the good fit 
 to the IR He I 10830\AA. Furthermore, in figure 20 the fit is extended beyond 1 micron. The good match to other
 He I-IR lines, adopting equal $v_{min}$(He I) and $\tau$(He I) as in the optical part of the spectrum, is clear
 and provides unambiguous evidence for the presence of helium in this event. 
  The OI 7773\AA~line with $\tau$=0.5, on the
 other hand, is found to be not so strong and 
 deep as is the case in most Type Ic events (Matheson et al. 2001). These two facts point to a Type Ib nature, more 
 than a Type Ic class. 
\begin{figure}
\vspace{0.3cm}
  \includegraphics[height=7cm,width=8.5cm]{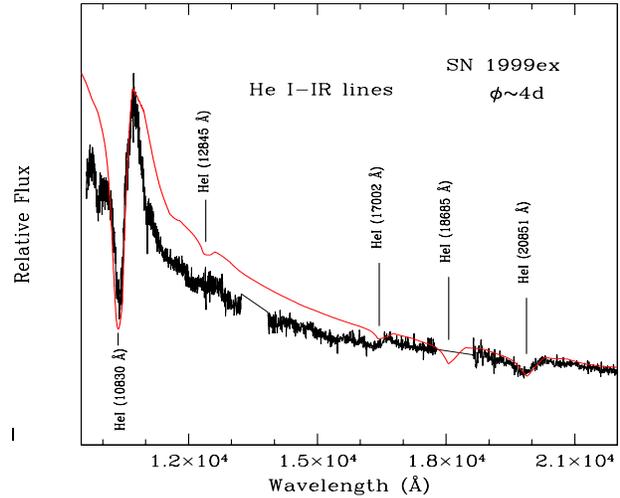}
\caption{The $~$4day IR spectrum of SN 1999ex compared with synthetic spectra that contain only lines of He I
 . The prominent features attributed to He I lines are indicated.}
\end{figure}
\begin{figure}
\vspace{0.3cm}
  \includegraphics[height=8.5cm,width=8.5cm]{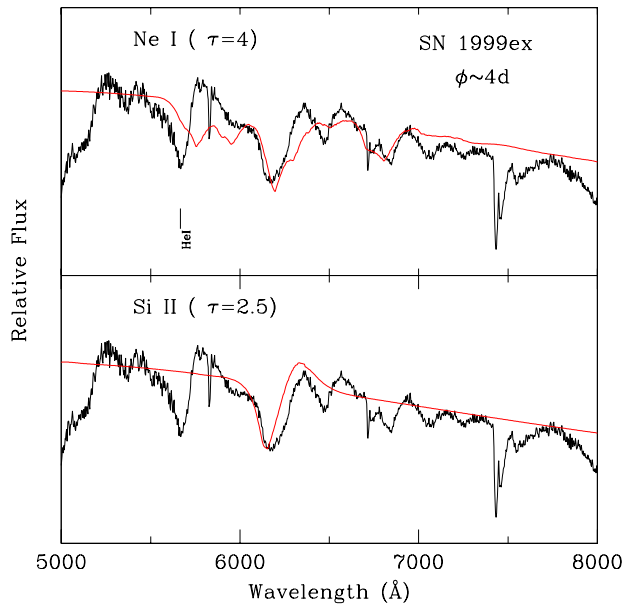}
\caption{The 4day spectrum of SN 1999ex is compared with synthetic spectra that contain only lines of Ne I
 (top panel) and Si II (bottom panel).}
\end{figure}

In the lower panel of Figure 19 we test the exponential case (see details in the ``SN 1984L'' part; Sect 3.1). We
 assign two components to the He I lines: one above $v_{min}$= 11000 km s$^{-1}$ with $v_e=$3000 km s$^{-1}$ and one 
 below $v_{min}$= 11000 km s$^{-1}$ with negative $v_e$ ($v_e=-$2000 km s$^{-1}$), such that $\tau$ is continuous 
 at the detachment velocity (i.e. at 11000 km s$^{-1}$). The fit is slightly improved for He I lines at 5876\AA~and 
 10830\AA~and also for the IR triplet Ca II.
\begin{figure}
  \includegraphics[height=9cm,width=9cm]{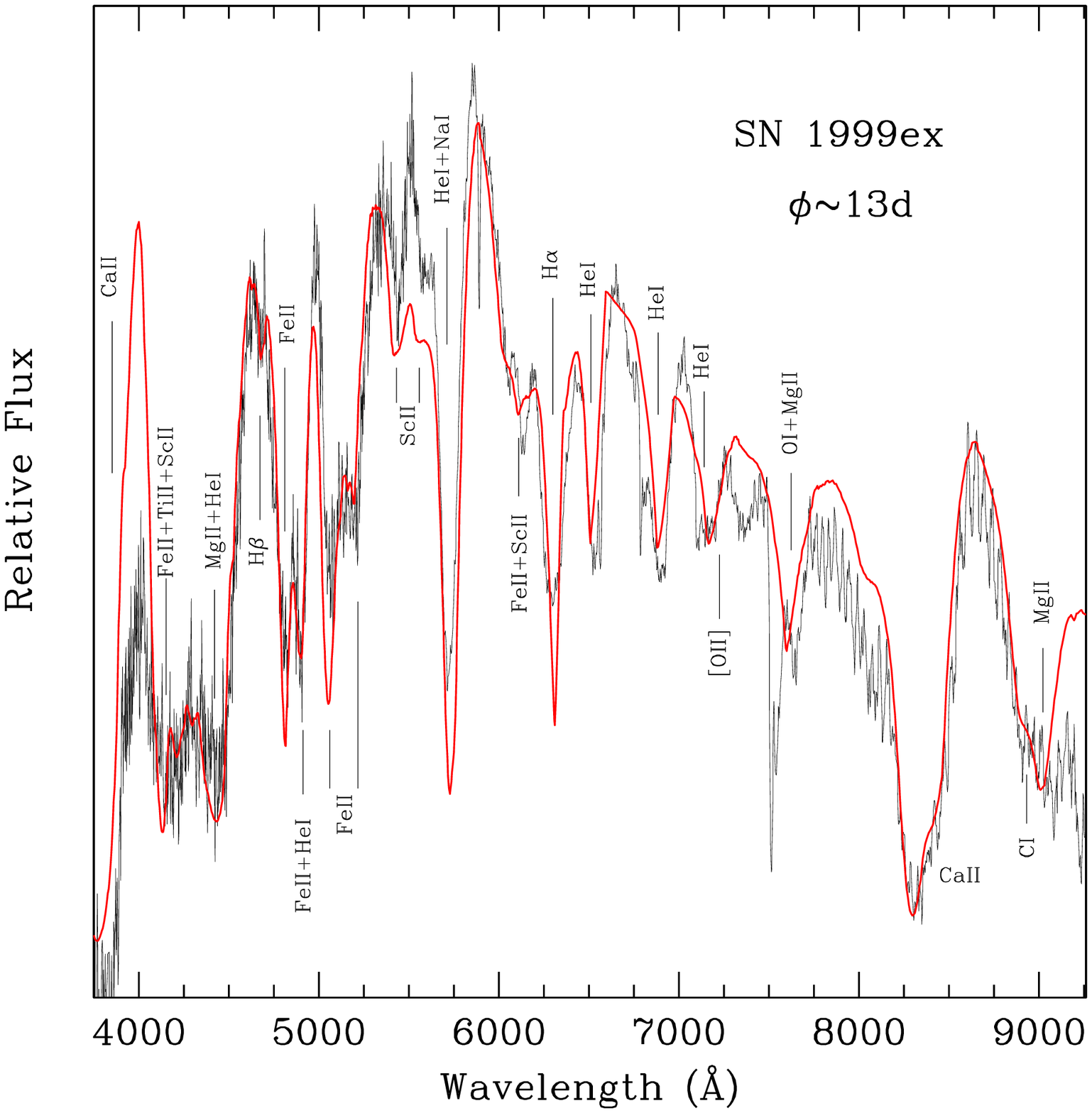}
\caption{SSp fit of SN 1999ex compared to the observed spectrum at day 13. Lines of conspicuous features are shown.}
\end{figure}

As far as the trough around 6250\AA~is concerned, we checked the Si II identification attributed by Hamuy et al.(2002).
 We test as well the Ne I possibility. The closer view in Figure 21 illustrates the two possibilities. In the upper 
 panel undetached Ne I lines can fit the feature, however they would introduce unwanted features in the rest of the 
spectrum. Furthermore, because of the depth of the observed trough, a good match with Ne I 6402\AA~would require an 
optical depth of 4, indicating a departure factor from ``LTE'' of $\sim$40 (Hatano et al. 1999), highly improbable 
  although non-thermal excitation of Ne I needs to be more thoroughly investigated (i.e. 
 through NLTE and hydrodynamic SSp codes). 
 The Si II possibility is shown in the bottom of Figure 21. The undetached Si II 6355\AA, with $\tau=$2.5, is rather blue 
to account for the feature. We note here that if one adopts C II 6580\AA, then one needs to assign it a very high 
velocity, about 8000 km s$^{-1}$ higher than the one of He I. The most likely identification therefore remains 
 H$\alpha$. In fact the best fit in Figure 19 is achieved using $v_{cont}$(H$\alpha$)= 8000 km s$^{-1}$ and 
 $\tau$(H$\alpha$)=1.45, while for day 13 H$\alpha$ has $v_{cont}$(H$\alpha$)= 5000 km s$^{-1}$ 
 and $\tau$(H$\alpha$)=1.6 (Fig. 22). The match at both
 phases is quite good. The other lines responsible for shaping the spectra are labeled in the figures. In addition to 
 the convincing match to lines of He I, IR-Ca II, O I, Fe II we note the presence of a weak absorption near 4680\AA~
that is well accounted for by H$\beta$ in our 13days SSp (Fig. 22). This can
 be taken as support  for the presence of hydrogen expelled at high velocities in this supernova.  

 It would have been of great interest to look at later spectra of this object, in order to see how the trough assigned 
 to H$\alpha$ would evolve with time.
\begin{figure}
\includegraphics[height=9cm,width=9cm]{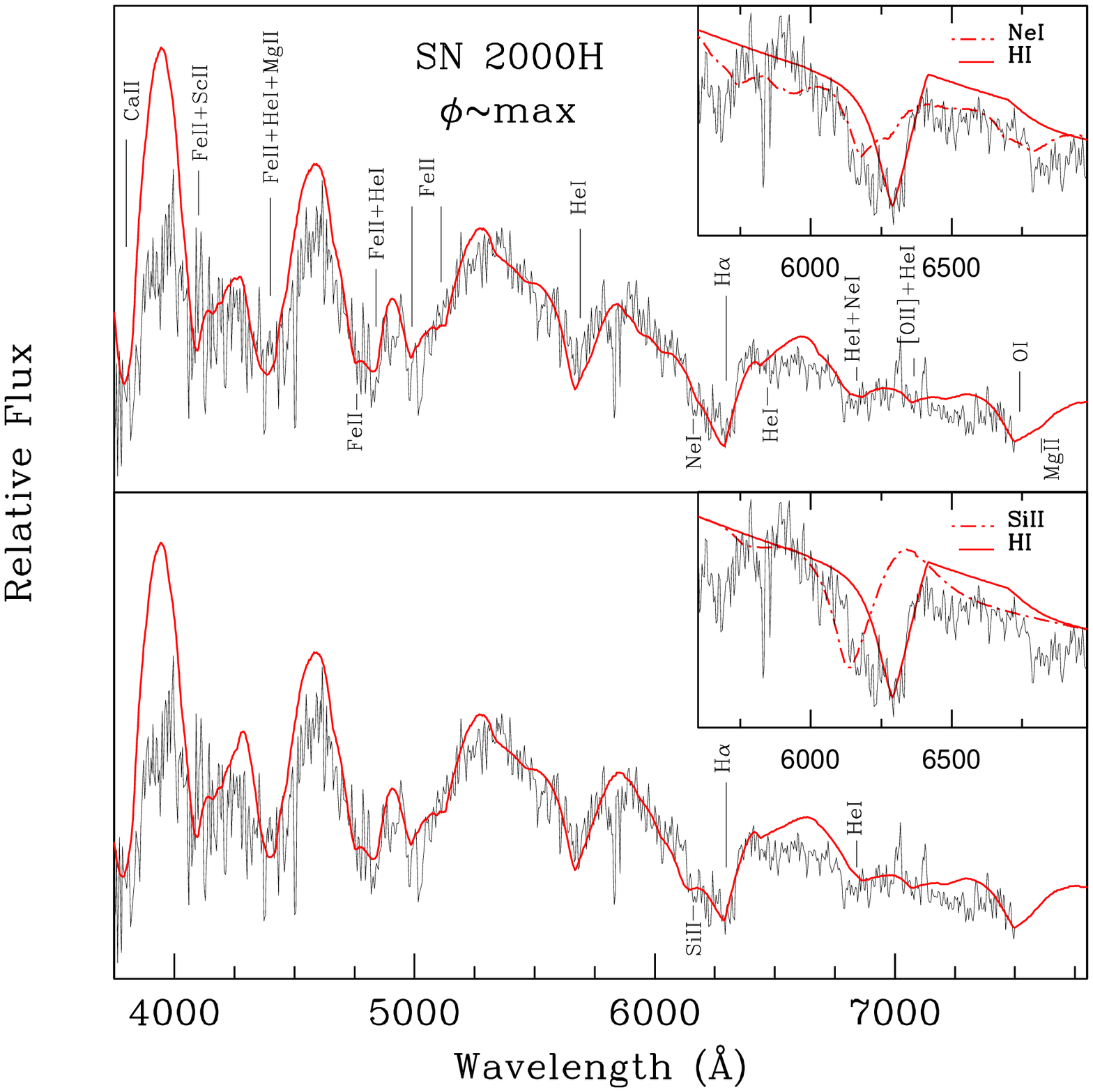}
\caption{SSp fit of SN 2000H compared to the observed spectrum at maximum light. Lines of conspicuous features 
 are indicated. The top panel shows the ``H$\alpha$+Ne I'' possibility for the broad feature near 6300\AA. A 
 corresponding zoomed view is displayed in the window. Similarly, the bottom panel shows the ``H$\alpha$+Si II'' case.}
\end{figure} 
\begin{figure}
\includegraphics[height=9cm,width=9cm]{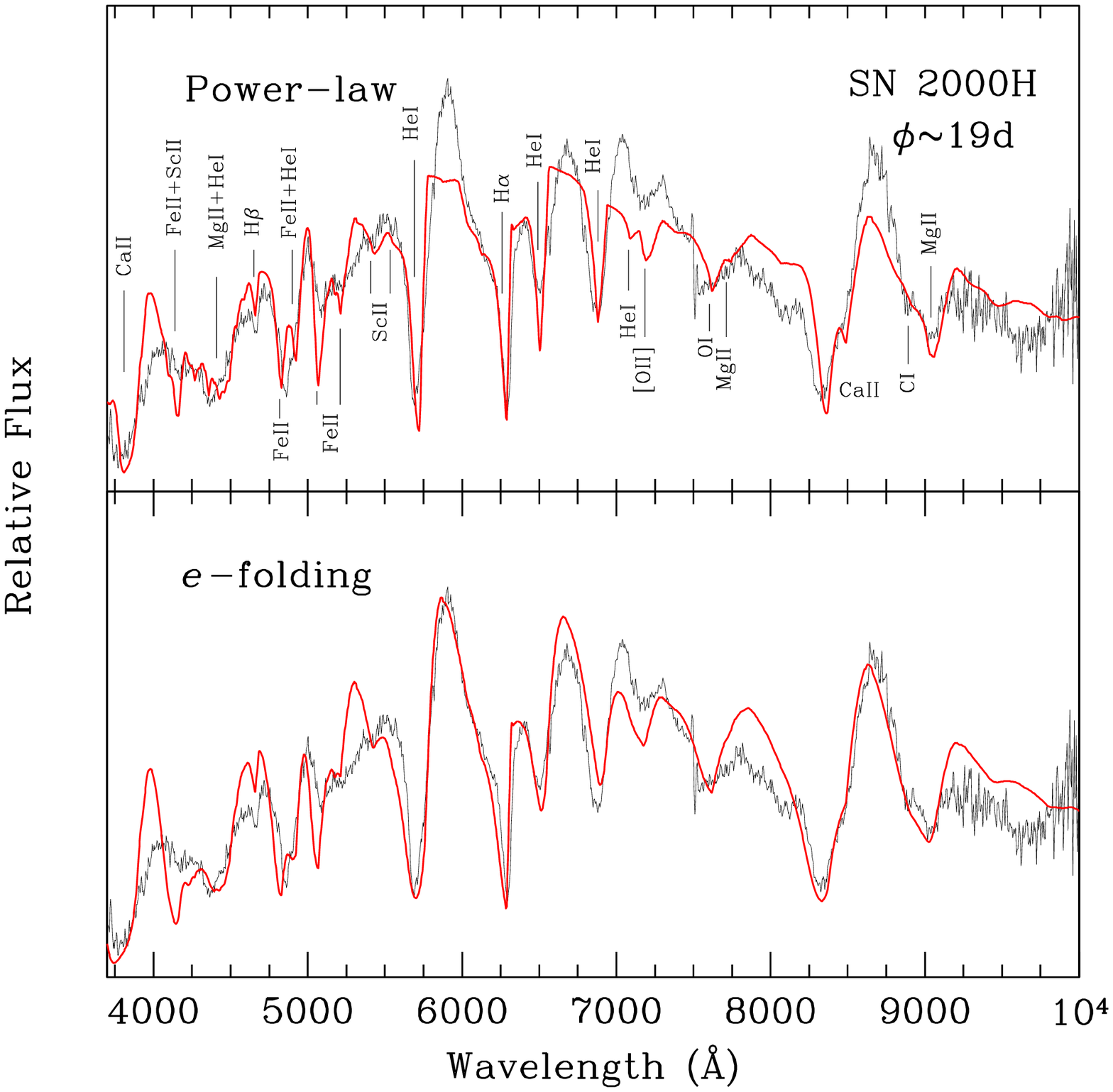}
\caption{SSp fit of SN 2000H compared to the observed spectrum at 19days. The lower panel shows the $e-$folding 
 optical depth possibility (see discussion). Conspicuous line features are shown. }
\end{figure}
\vspace{0.4cm}\\
$*$ {\bf{SN 2000H:}}

The Type Ib SN 2000H is considered one of the more interesting Ib-c objects, especially with a strong and deep trough 
near  6300\AA~(Benetti et al. 2000; Branch et al. 2002). Here we analyze three spectra at different epochs, namely at 
 maximum light, 19days and 30days. Figure 23 compares the observed spectrum around maximum with the resulting best fit 
SSp which has $v_{phot} = 11000$ km s$^{-1}$ and  T$_{bb}=$10000 K. The overall match is quite good. Undetached He I 
 lines, with $\tau =$2, provide a good fit with the He I 5876\AA, while features assigned to He I lines 6678\AA~and 
 7065\AA~ are very weak. The absorption 
trough near 6280\AA~is exceptionally broad and cannot be accounted for only by H$\alpha$. We identify two possibilities 
 for which we obtain a broad absorption in agreement with the observed feature. The upper panel in Figure 23 shows the 
 ``H$\alpha$+Ne I'' combination, while in the lower panel the ``H$\alpha$+Si II'' combination is illustrated. In both 
 panels a closer view of the 6280\AA~region 
is displayed in the window. In the first case, undetached NeI with $\tau =$1.5, a contribution blueward of
 He I 7065\AA~improves the fit with the observed feature, however the He I 5876\AA~emission component is 
 under-estimated. Note that the narrow absorption in the emission peak is attributed to Na ID interstellar line 
 originating in the parent galaxy. In the  lower panel, undetached 
Si II ($\tau =$3) improves the match with the observed absorption trough without altering  the resulting SSp at the 
 emission part of the He I 5876\AA~P-Cygni profile. As a result we believe that the``H$\alpha$+Si II'' combination 
is the most probable.   

 Figures 24 and 25 compare spectra at 19 and 30 days with their corresponding best fit SSp. 
  These have $v_{phot} = 6000$ km s$^{-1}$ and  T$_{bb}=$4600 K (Fig. 24) and  $v_{phot} = 5000$ km s$^{-1}$ 
 and  T$_{bb}=$6000 K (Fig. 25). The two spectra are similar in having narrower features than the maximum light 
 spectrum, resulting from observing at smaller radii where expulsion velocities are lower. Lines of 
 Ca II, [O II], O I, Mg II are seen to develop.  
 Introducing Sc II lines helps to form a feature blueward of the strong He I 5876\AA~P-Cygni profile. A distinct 
 feature appears redward of the IR Ca II profile, more evident in the 30day spectrum, and is accounted for by 
 C I ($\tau=$0.2 at 19days and $\tau=$0.8 at 30days). The noticeable change compared to the spectrum at maximum is 
 the development of the He I 6678\AA~and 7065\AA~troughs. The He I reference line has 
 $v_{cont}$(He I)= 2000 km s$^{-1}$ and $\tau$(He I)=5 
 at day 19, while at day 30 the corresponding He I parameters are $v_{cont}$(He I)= 2000 km s$^{-1}$ and 
 $\tau$(He I)=3.4.  
\begin{figure}
\includegraphics[height=9cm,width=9cm]{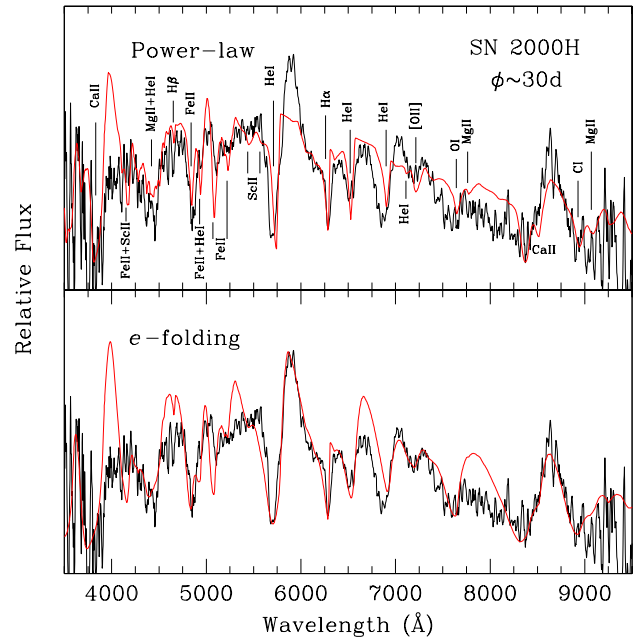}
\caption{SSp fit of SN 2000H compared to the observed spectrum at 30days. The lower panel shows the $e-$folding 
 optical depth possibility. Conspicuous line features are indicated. }
\end{figure}

 Similarly to what is seen in the case of SN 1984L, the He I profiles retain rounded emission components that cannot 
 be matched by a power-law SSp. We tried to improve the He I fits by switching to the $e-$folding assumption for the 
 optical depth, introducing a continuous two-component behaviour of the He I optical depth: one above 
 $v_{min}$= 8000 km s$^{-1}$ with $v_e=$3000 km s$^{-1}$ and a second component with negative $v_e$
  ($v_e=-$2000 km s$^{-1}$), such that $\tau$ is continuous at the detachment velocity (i.e. at 8000 km s$^{-1}$ 
 for the 19day spectrum and 7000 km s$^{-1}$ for the 30days one). It is important 
 to recall here that the line should not be said to be ``detached''. It has however a maximum value of $\tau$ that 
 is not at the photosphere as is ordinarily the case for an undetached line.
The bottom panels in Figures 24 and 25 illustrate the resulting synthetic spectra in the $e-$folding cases. The fit 
 is somewhat improved compared to the power-law case. The noticeable improvements concern in particular the He I 
 features and the infra-red Ca II profile.
 We note the good match to the $\sim$6250\AA~trough with H$\alpha$. The best fit is achieved using  
 $v_{cont}$(H$\alpha$)= 7000 km s$^{-1}$ and $\tau$(H$\alpha$)=2.5  at day 19 (Fig. 24), while for day 30 
 H$\alpha$ has $v_{cont}$(H$\alpha$)= 8000 km s$^{-1}$ and $\tau$(H$\alpha$)=1.45 (Fig. 25). No alternative 
 identification to H$\alpha$, that would be logically acceptable, has been found. In addition, the H$\alpha$ 
 identification is strongly supported by the presence of the absorption notch near 4660\AA, well accounted for 
 by H$\beta$ in our resulting synthetic spectra. 
\subsection{Type Ic SNe: the representatives}
$*$ {\bf{SN 1987M:}}
\begin{figure}
  \includegraphics[height=9cm,width=9cm]{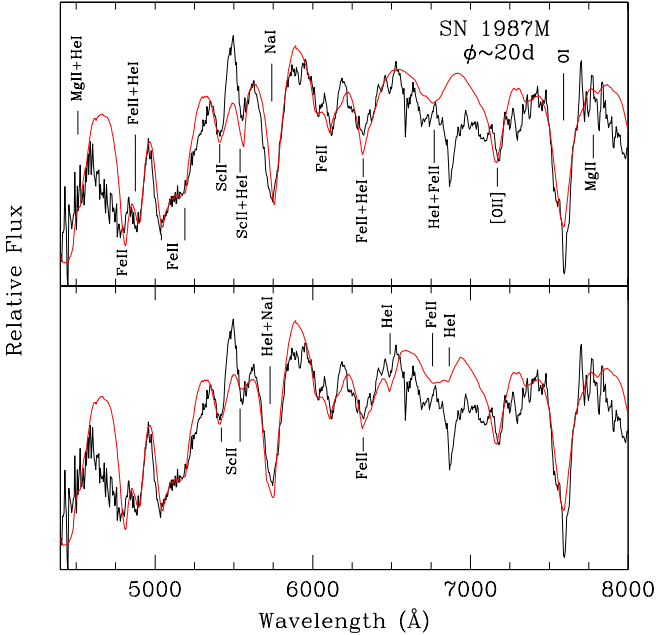}
\caption{SSp fit of SN 1987M compared to the observed spectrum at 20days.
 The upper panel shows the high He I velocity case, while the bottom panel displays the low 
 He I velocity case (see discussion). Conspicuous line features are shown. }
\end{figure}                                      

In Figure 26 the observed spectrum of SN Ic 1987M, near day 20, is compared to the generated SSp with 
 $v_{phot}=7000$ km s$^{-1}$ and  T$_{bb}=$4500 K. The OI 7773\AA~trough is clearly deep and is well matched with 
 our SSp ($\tau=$5). Identifying He I lines in this object is problematic and was a subject of different 
 investigations (Jeffery et al. 1991\nocite{Jeff91}; Clocchiatti et al. 1996b). In fact, Jeffery et al. (1991) 
 have presented a 
 synthetic-spectrum analysis around maximum and came to the conclusion that He I lines may be present. In addition 
 they also claimed the presence of a weak H$\alpha$ trough. In our SSp-LTE 
 approach applied to the 20day spectrum, we tested two possibilities for He I identification, namely He I lines 
 at high velocity ($v_{cont}$(He I)= 10000 km s$^{-1}$; $\tau$(He I)=0.16) and a lower velocity case 
 ($v_{cont}$(He I)= 2000 km s$^{-1}$; $\tau$(He I)=0.6). The upper panel in Figure 26 corresponds to the first case, 
 in which the strong P-Cygni profile at $\sim$5750\AA~is due to Na I line alone. The high velocity He I 5876\AA~then 
 contributes to the trough blueward of the Na I profile. The undetached Sc II lines are in part responsible for the 
 two absorption features indicated in the plot ($\tau$(ScII)=0.7). The
 other He I optical lines, at 6678\AA\ and 
 7056\AA, both contribute to the strong Fe II features ($\tau$(FeII)=22). In the low velocity case (Fig. 26; 
 lower panel) a weak He I absorption at $\sim$6500\AA~appears, which provides a better match to the observed spectrum,
  as does ``He I+Na I D'' at $\sim$5750\AA. It is not simple to decide which fit to adopt from analyzing only one  
 spectrum. The two possibilities seem plausible and both indicate the presence of He I lines in SN 1987M at the 
 age of 20 days.

 We find no need to include either Si II and/or C II or H$\alpha$. We note however that Si I lines may help to fit 
 the deep trough around 6900\AA, but they would introduce then various unwanted features in 
 the rest of the spectrum.   
\vspace{0.4cm}\\
$*$ {\bf{SN 1994I:}}
\begin{figure}
  \includegraphics[height=9cm,width=9cm]{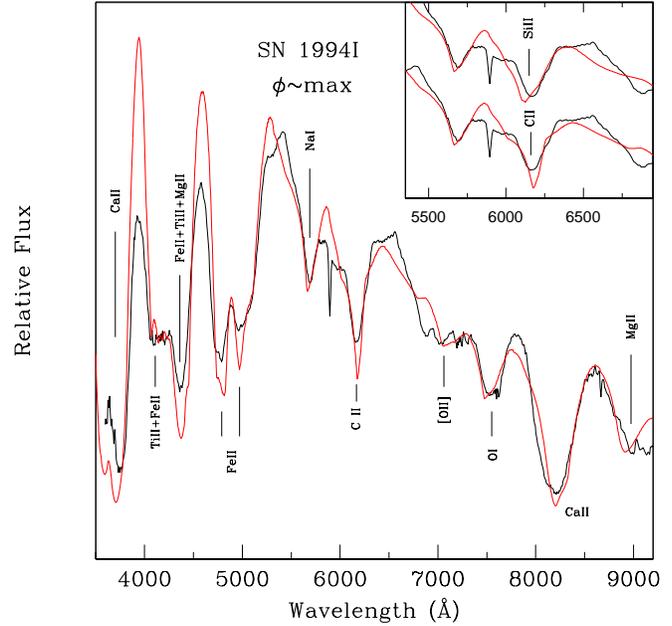}
\caption{SSp fit of SN 1994I compared to the observed spectrum at maximum brightness. Conspicuous line 
 features are shown. The window shows the C II and Si II identification cases  (see discussion).}
\end{figure}                                      

SN 1994I is considered the best observed ``normal'' Type Ic event. Various works have presented detailed analysis 
of line identifications in this object (Wheeler et al. 1994; Filippenko et al. 1995\nocite{Fil95}; 
 Clocchiatti et al. 1996b; Millard et al. 1999\nocite{Mill99}). The most controversial issues 
 concerned traces of H and He in the early spectra. Filippenko et 
al. (1995) have attributed the observed absorption feature near 10250\AA~to the infrared He I 10830\AA, arguing 
for the presence of helium in the ejecta of SN 1994I. Millard et al. 1999, however, clarified
 the related issues by means of synthetic spectra. One of the most important results was the incompatibility  of the
 simultaneous fit of the $\sim$10250\AA~observed feature with He I
 10830\AA\ and with the optical He I lines. Contrary 
to the case of SN 1999ex (see Fig. 19), trying to account for the observed absorption near 10250\AA~with 
He I 10830\AA~introduces too strong He I features in
  the optical region (Millard et al. 1999). Even with a NLTE synthetic spectra analysis, Baron et al. (1999)\nocite{Bar99} 
 were not able to reproduce the $\sim$10250\AA~infrared feature assuming it to be the He I 10830\AA~line. Instead, 
 the feature in question may be accounted for by lines of C I and/or Si I (Baron et al. 1999; Millard et al. 1999).   

 Figure 27 compares the observed spectrum around maximum with an SSp that has $v_{phot} = 12000$ km s$^{-1}$ and  
 T$_{bb}=$7000 K. The match is quite good. The trough near 5700\AA~is accounted for by the undetached Na ID 
 line ($\tau=1$). The other prominent features are well described by lines of Fe II($\tau=10$), 
 Ca II($\tau=300$), O I($\tau=0.7$), [O II]($\tau=0.2$). Lines of Mg II, 
 $\tau=1$, have been introduced to help the fit blueward the IR-Ca II P-Cygni profile. Lines of Ti II, 
 $\tau=0.8$, are also considered to help the fit in the 4000$-$4500\AA~region.  

  Concerning the feature near 6180\AA, it was difficult at some epochs in the spectra presented by Millard et al. 
 (1999), to decide between detached C II 6580\AA~and undetached Si II 6355\AA. Moreover, at -2days and -4days, the 
 authors obtained a better fit with a combination of the two. In the spectrum 
 displayed in Figure 27 we test the two possibilities, namely undetached Si II ($\tau=1.5$) and detached C II 
 ($v_{cont}$= 8000 km s$^{-1}$; $\tau$=1.7$\times$10$^{-3}$). A closer view of the region of interest is 
 displayed in the window. We found Si II somewhat too blue to account for the observed feature. There is no need 
 at this epoch to combine the two lines. Therefore we prefer detached C II as the more probable. 
\vspace{0.4cm}\\
$*$ {\bf{SN 1996aq:}}
\begin{figure}
  \includegraphics[height=9cm,width=9cm]{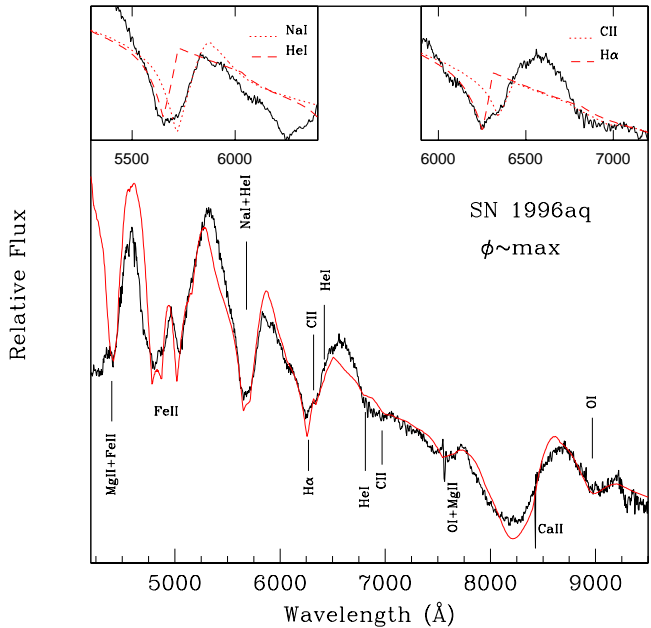}
\caption{SSp fit of SN 19996aq compared to the observed spectrum at maximum light. The right and left windows 
 show a zoomed view of the 6300\AA~and 5700\AA~regions, respectively. Conspicuous line features are also reported.}
\end{figure}                                      
 
 We analyzed two photospheric spectra of SN 1996aq, namely one near maximum light and one at $\sim$24days. The 
 supernova has been classified as Type Ic SN near maximum when first observed (Nakano et al. 
 1996\nocite{Nak96}). The observed spectrum near maximum is compared, in Figure 28, with a 
 synthetic spectrum that 
 has  $v_{phot} = 9000$ km s$^{-1}$ and  T$_{bb}=$9000 K. The SSp accounts for almost all the conspicuous features, 
 namely IR-Ca II, Fe II, O I and Mg II. The trough near 5680\AA~appears broader than usual. To account for this 
 absorption feature we use a combination of undetached Na ID ($\tau=1$) and detached He I ($v_{cont}$= 
 3000 km s$^{-1}$; $\tau=0.35$). The resulting profile fits nicely the broad trough (left window in Fig. 28). 
 The absorption features due to the He I lines 6678\AA~and 7065\AA~
 are too weak to be clearly seen as indicated in Figure 28. Concerning the 
 $\sim$6300\AA~feature, while Si II is found to be too blue, we find that a combination of C II ($v_{cont}$= 2500 
 km s$^{-1}$; $\tau=2.2 \times 10^{-3}$) and H$\alpha$ ($v_{cont}$=
 6000 km s$^{-1}$; $\tau=0.12$) provide 
 a satisfactory fit. On the one hand, the weak feature near 6330\AA~is produced by the minimum of the C II 6580\AA~
 absorption. This possibility is illustrated in the right window of Fig. 28. On the other hand, a weak absorption 
 in the emission feature near 4560\AA~could be due to H$\beta$ although the H$\alpha$ optical depth seems 
 too small for H$\beta$ to be plausible.

 Alternatively, if we try to fit the 6300\AA~feature by only C II, then this latter should be expelled at very high 
 velocity, even more than helium. Moreover, in this way we could not produce the notch near 6330\AA. We regard the 
 ``C II+H$\alpha$'' combination as the more probable. However 
 a final confirmation of this is still not beyond doubt. The identification of H$\alpha$ in Type Ic  was a subject 
 of different discussions (Jeffery et al. 1991; Filippenko 1992\nocite{Fil92}; Swartz et al. 1993;  
 Wheeler et al. 1994; Branch 2003). Generally 
 the identification of H$\alpha$ in early spectra of some SNe Ic objects (exp. SN 1987M and SN 1994I) has not been 
 accepted. In SN 1987M for example the measured 
 velocity for H$\alpha$ was suspiciously low compared to calcium and
 oxygen. In the case of SN 1996aq we find a different
 situation, since the expansion velocity attributed to H$\alpha$ is the highest (i.e. higher
  than C II and He I while the other lines are all undetached). It would be interesting to have more detail for this 
 object and to have a larger sample of early spectra of Type Ic objects.     

 Figure 29 compares the 24day spectrum to the SSp that has $v_{phot} = 4600$ km s$^{-1}$ and  T$_{bb}=$5000 K. Ions 
 that are responsible for the most
 conspicuous supernova absorption features are indicated. The match is good. Note both the strong Ca II infrared 
 triplet and Ca II H$\&$K are well produced in the SSp. Ni II helps the fit longward of the Ca II H$\&$K profile. The 
 double absorption feature near 9000\AA~is well produced by undetached lines of C I($\tau=1$) and Mg II($\tau=3$). 
 Lines of O I and [O II](at 7321\AA) are also introduced to account for features near 7640\AA~and 
 7220\AA, respectively. A noticeable feature of the SN at this phase is the clear appearance of He I lines, that question the 
 classification of SN 1996aq as Type Ic object. The broad P-Cygni profile with the absorption minimum near 5720\AA~
 cannot be accounted for by Na ID alone. We favour He I 5876\AA~ because
 it implies other observed features,
 especially the He I lines at 6678\AA~ and 7065\AA. He I lines in the SSp are produced adopting 
 $v_{cont}$= 2400 km s$^{-1}$ and $\tau=0.25$. 

\begin{figure}
  \includegraphics[height=10cm,width=9cm]{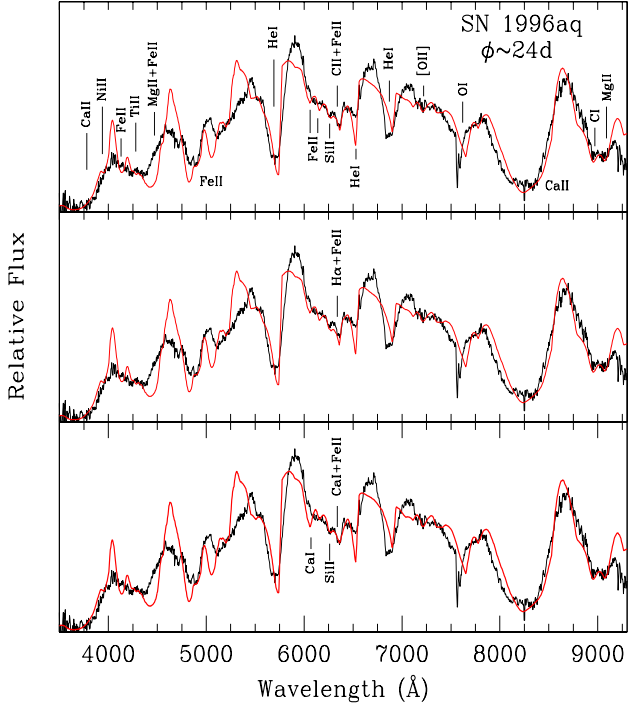}
\caption{SSp fit of SN 1996aq compared to the observed spectrum around day 24.
 The different panels show the possible line identifications for the 6300\AA~absorption, namely C II 
 (upper panel), H$\alpha$ (middle panel) and Ca I (bottom panel). Conspicuous line features are reported. }
\end{figure}                                      
 The slope in the 5950$-$6450\AA~wavelength range does not contain conspicuous features although it has some weak 
 absorptions, reminiscent of what is seen for example in SNe 1984L and 1998T at similar phases. We regard SN 1996aq 
 as a transition object between Ib and Ic, rather than a $''pure''$ Type Ic event. We have checked different line 
 combinations in order to decide which ions to introduce to fit the 5950$-$6450\AA~region, and especially the feature 
 near 6355\AA. The different panels in Figure 29 illustrate these possibilities. In fact, while lines of Fe II and 
 Si II(undetached) provide a good fit to the observed features, we find it difficult to decide between C II, 
 H$\alpha$ and Ca I to account for the absorption near 6355\AA. The three cases correspond respectively to: 
 C II with $v_{cont}$= 5400 km s$^{-1}$ and $\tau=7 \times 10^{-6}$ (upper panel); H$\alpha$  with 
 $v_{cont}$= 4900 km s$^{-1}$ and $\tau=0.02$ (middle panel); undetached Ca I with $\tau=6$ (lower panel). The Ca I 
 case, lower panel, produces an unwanted absorption trough near 6060\AA. For the C II case, upper panel, the element 
 is assigned a velocity 3000 km s$^{-1}$ greater than the one of helium, which is hard to accept. Therefore the most 
 probable situation remains the case of the high velocity hydrogen (middle panel).
\subsection{Type IIb SNe: the representative}
$*$ {\bf{SN 1993J:}}

 The discovery of the $''hybrid''$ Type IIb SN 1993J has created a link between SN Ib-c and Type II objects, 
 creating and extending our understanding of the physics of core-collapse. At early phases, this SN displays 
 conspicuous hydrogen Balmer features similar to Type II SNe. At the nebular phase however, the spectrum shows many 
 signatures of Type Ib-c SNe, namely strong oxygen and calcium lines (Filippenko et al. 1994\nocite{Fil94}; 
 Lewis et al. 1994\nocite{Lew94}; Matheson et al. 2000\nocite{Math00}). Furthermore, the analysis 
 of late epoch spectra reveals that traces of hydrogen (i.e. H$\alpha$ in emission were still present 
 (Patat et al. 1995\nocite{Pat95}). Another well observed SN belonging to Type IIb class 
 is SN 1996cb. By comparison with SN 1993J, SN 1996cb showed Balmer lines with stronger P-Cygni profiles. In addition, 
 the photosphere of SN 1996cb receded faster than its counterpart in SN 1993J and the He I features (in absorption) 
 appeared earlier, around day 24 in SN 1993J and near maximum light in SN 1996cb 
 (Qiu et al. 1999\nocite{Qiu99}; Deng et al. 2001\nocite{Den01}).  

\begin{figure}
  \includegraphics[height=10cm,width=9cm]{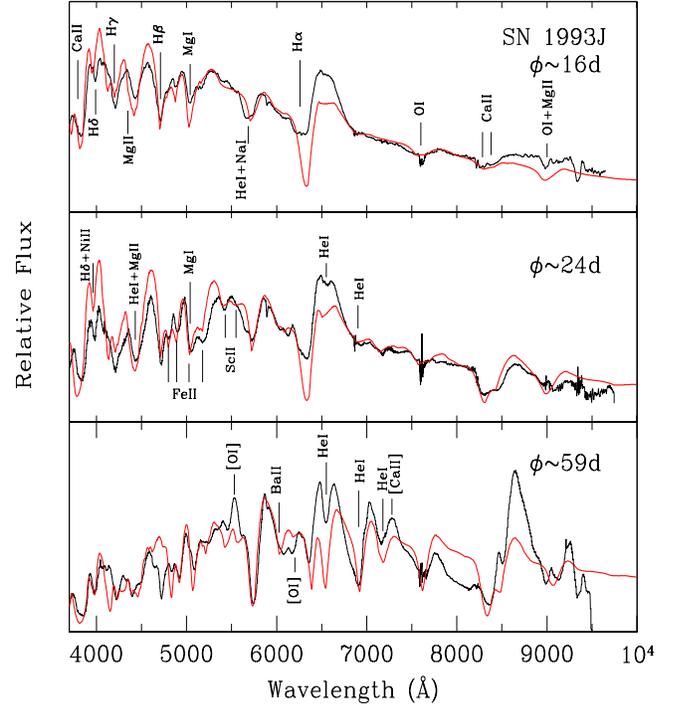}
\caption{ SSp fit of SN 1993J compared to the observed spectra at 16days (upper panel), at 
 24days (middle panel) and at 59days (lower panel). Conspicuous line features are reported.}
\end{figure}                                      
 We have analyzed spectra at three different epochs for SN 1993J, namely at 16, 24 and 59~days. Figure 30 
 displays our best fit spectra compared to the observe ones. The plotted synthetic spectra have 
 $v_{phot} = 9000$ km s$^{-1}$ and  T$_{bb}=$7800 K (16days; upper panel),  $v_{phot} = 8000$ km s$^{-1}$ and  
 T$_{bb}=$7000 K (24days; middle panel) and  $v_{phot} = 6000$ km s$^{-1}$ and  T$_{bb}=$5000 K (59days; 
 lower panel). The fit to the 16day spectrum is good, except the H$\alpha$ profile. The corresponding strong and 
 broad P-Cygni feature cannot be produced completely by the ``SYNOW'' code. This is because in our SSp 
 treatment we are adopting a resonance scattering source function. The hydrogen reference line has 
 $v_{cont}$= 1000 km s$^{-1}$, while an optical depth of $\tau=21$ is used. Profiles of H$\beta$, H$\gamma$ and 
 H$\delta$ are clearly discernible. The P-Cygni profile around 5700\AA~is assigned to He I 5876\AA, with a small 
 contribution from Na ID. The other optical He I lines are not visible at this phase. The helium is found to be 
 undetached and has a moderate optical depth of 0.6. The shallow IR-Ca II triplet 
 is fitted adopting $\tau=20$. The observed Ca II H$\&$K is nicely reproduced in the SSp. 

 As the supernova evolves,
  the IR-Ca II profile becomes clearer (Fig. 30; middle panel). At this phase an optical depth of $\tau(CaII)=300$ 
 is used. Fe II lines are also clearer compared to the 16day spectrum ($\tau(FeII)=3$ at 16days and 
 $\tau(FeII)=12$ at 24days). The notch appearing on the emission component of H$\alpha$ is attributed to He I 6678\AA. 
 Helium is still undetached with an optical depth of 1.5.
 SN 1996cb at a similar phase, near 25days, has more evident He I lines compared to SN 1993J, especially the lines 
 at 6678 and 7056 (Fig 2; Deng et al. 2001). In fact, its 25day spectrum resembles the 59day spectrum of SN 1993J 
 rather than the 24days one. The He I lines at day 59 are prominent (Fig. 30; bottom panel). The He I reference line is 
 undetached and has $\tau(He I)=30$. The corresponding SSp fit is quite good for the 5876\AA~line, while it is 
 difficult to obtain a good match with the other optical He I lines. 

 Hydrogen Balmer lines become weak at this phase. Ba II lines 
 are introduced to improve the fit just blueward of He I 5876\AA~($\tau(BaII)=5$). At this phase, forbidden emission 
 lines develop, indicating the transition to the nebular phase (exp. [O I] 5577\AA~, [O I] 6300,64\AA~and 
 [Ca II] 7291,7324\AA; bottom panel). The O I 7773\AA~absorption profile is reproduced at the three different 
 epochs by $\tau=0.3$(at 16days), $\tau=0.4$(at 24days) and $\tau=2.5$(59days). Ni II lines are also introduced in 
 the 24days and 59days synthetic spectra, blended with H$\delta$, to improve the match with the observed weak 
 absorption on the Ca II H$\&$K emission component.
 
\subsection{Type II SNe: the representative}
$*$ {\bf{SN 1999em:}}

 As a representative of the Type II class, we have analyzed SN 1999em. The well sampled observations, spectra and 
 photometry, have been presented and studied by Elmhamdi et al. (2003\nocite{Elm03}). The event shows 
 characteristics typical of 
 a Type II-P object, namely clear and broad Balmer P-Cygni profiles and
 a plateau phase of almost constant luminosity in the optical. 

 Figure 31 displays two spectra during the photospheric phase. For comparison with other events of our CCSNe sample, 
 the phases are normalized to maximum light in Type Ib-c SNe, adopting 16 days as the rise time to reach maximum. The 
 observed spectra are compared to the best fit synthetic ones that have $v_{phot} = 10000$ km s$^{-1}$ and 
 T$_{bb}=$10000 K (-6days; upper panel) and  $v_{phot} = 4600$ km s$^{-1}$ and  T$_{bb}=$6400 K (25days; lower 
 panel). Synthetic line features are designated according to the ion whose 
 line gives rise to the feature (Fig. 31). At the earliest phase, i.e. -6days, only three elements are introduced 
 in the SSp model. Indeed undetached lines of Balmer hydrogen, He I and a weak contribution from Ca II are almost 
 sufficient to reproduce the most conspicuous features superimposed on the ``hot'' continuum. The fit with Na I D, 
 at this phase, is poor compared to He I 5876\AA. Moreover He I contribution, in Type II SNe, is found to be 
 important shortly after the explosion. In fact in SN 1987A, the He~I 5876\AA~feature was clearly present during 
 the first few days, then owing to the decreasing envelope-temperature it rapidly faded and disappeared completely 
 around 1 week after the explosion, when Na I D starts to emerge (Hanuschik $\&$ Dachs, 1988)\nocite{Han88}.
 Baron et al. (2000), in analyzing very early spectra of SN 1999em, have found evidence for helium enhanced by  
 at least a factor of 2 over the solar value. The interpretation of the He I strength in terms of helium-overabundance
 is however premature, since freeze-out effects were not considered and could lead to the enhanced helium-excitation 
 compared to the steady-state model (Utrobin \& Chugai 2002)\nocite{Ut02}. 
 
\begin{figure}
  \includegraphics[height=9cm,width=9cm]{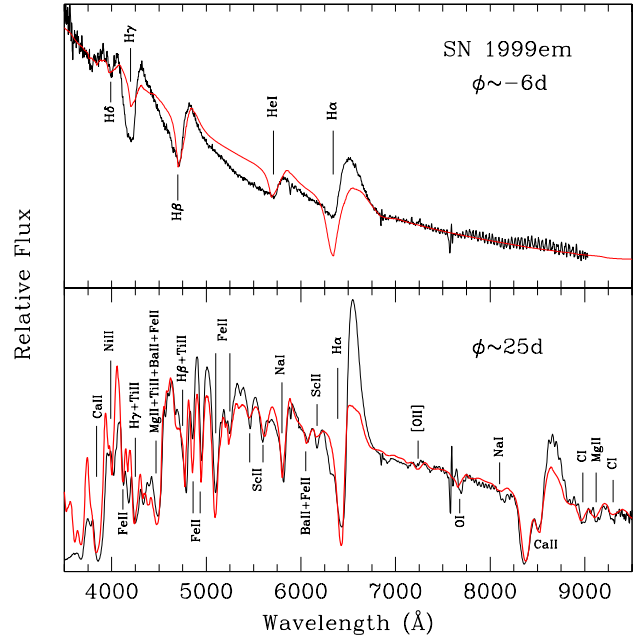}
\caption{SSp fit of SN 1999em compared to the observed spectra at day -6 (upper panel) and at day 25 
 (lower panel). Conspicuous line features are shown. For comparison, the reported phases 
 are normalized to maximum in Type Ib-c SNe,  using ``16 days'' as the rise time.}
\end{figure}                   
 Balmer hydrogen features, with $\tau=15$, are clearly evident. As in the
 case of SN 1993J, we cannot reproduce the full observed H$\alpha$ profile for the same reason. At day 25, H$\alpha$ 
 P-Cygni profile becomes narrower because of the decrease in the expansion velocity (Fig. 31; bottom panel). 
 H$\alpha$ in the SSp has now $v_{cont}$= 1000 km s$^{-1}$ and  $\tau=21$. The envelope
 temperature decreases and many lines emerge at this phase. Apart from the hydrogen lines (slightly detached), all the 
 lines are undetached. The match with almost all observed features is good. The Ca II H$\&$K and IR-Ca II are nicely 
 reproduced. A few distinct absorption lines redward of the near infrared Ca II triplet are identified as  C I, 
 $\tau=0.6$, through our spectral synthesis. Sc II lines, $\tau=3$, account rather well for features near 5450, 5600 
 and 6170\AA. The blue part of the spectrum, $\lambda < 5000$\AA, is subject to severe line blending, but it seems 
 well matched with contributions from lines of Fe II($\tau=10$), Ni II($\tau=5$), Ti II($\tau=8$), Ba II($\tau=2$), 
 Mg II($\tau=2$), Ca II and hydrogen. The feature near 7670\AA~ is attributed to O~I 7773\AA($\tau=0.3$), 
 while there may be a contribution from [O II]  around 7230\AA. 
\section{Spectroscopic mass estimates}
In the following we adopt two methods in order to obtain rough estimates of some ejecta masses (i.e. the total 
ejecta mass and/or a given element mass). The methods use the results from our spectral best-fits.

\subsection{Method 1}
 We compare the optical depths derived from our best-fits to the optical 
 depth plots for different compositions presented in the work by Hatano et al.(1999), making sure of course that the 
 same reference lines are used. In that paper the authors presented a systematic survey of ions that could be 
 responsible for supernova features in six different compositions. The LTE optical depth of each reference line, of 
 a given ion, is computed and then plotted against temperature. We focus on the ratio of hydrogen to helium optical 
 depths, i.e $\tau$(H I)/$\tau$(He I), attempting thereafter to translate this ratio into a relative abundance of the 
 two elements. This is important as it can give us a rough idea about the amount of hydrogen present in these 
 objects. The optical depth ratio is assumed to trace the abundance ratio of the two elements. This is
 essentially due to the fact that the $\tau$-calculations of Hatano et al. (1999) adopted a fixed
 electron density. The $\tau$(H I)/$\tau$(He I) ratio, indeed, depends on electron density, temperature
 and hydrogen to helium abundance ratio. Therefore for a given temperature, the $\tau$(H I)/$\tau$(He I)
 ratio depends  in principle only on the abundance ratio.
  We must remember however that the ``$\tau$-plots'' are for LTE, with no allowance
  for non-thermal excitation is made. For helium, for example, non-thermal excitation effects are significant in developing 
 He I features (Lucy 1991). More detailed estimates might take this effect into account, if one has a clearer
 understanding of outward mixing of radioactive material.

 For SN 1990I around maximum and according to our best SSp fit, the optical depth ratio, evaluated
 at the same velocity\footnote{The optical depths reported in Table 2 are evaluated at the detachment 
 velocity of the corresponding reference lines (e.g. for SN 1990I at maximum $\tau$(H I) at 
 16000 km s$^{-1}$and $\tau$(He I) at 14000 km s$^{-1}$).
 For a ``$v^{-8}$'' power-law, the optical depth of He I evaluated at the same velocity of H I is then:\\ 
 $\tau$(HeI)= 2.9$\times$(14/16)$^{8} \simeq$ 1.0},
 is measured to be 
 $\tau$(H I)/$\tau$(He I)$\sim$0.6. At similar high temperatures (i.e. $\sim$14000 K), one obtains a value of about 
 40 for that ratio according to the optical depth plots of different ion reference lines for
 a hydrogen-rich composition in Hatano et al. (1999) (see their Fig. 2). In addition, in the cited work the 
 $\tau$(H I)/$\tau$(He I) ratio remains 
 almost constant at high temperatures, then it starts increasing once the temperature decreases. For example 
 around T$\sim$10000 K, the ratio is  about 80. Near 9000 K, however the ratio reaches a value of about 1000. 
 These values (i.e. 40, 80 and 1000 at their corresponding temperatures) 
 correspond to a factor of 10.23 in abundances between hydrogen and helium (Table 1; Hatano et al. 1999). 
 Assuming proportionality, we obtain then a rough estimate of the hydrogen to helium relative-abundance for the 
 case of SN 1990I around maximum. In fact a relative abundance of $\sim$0.15 is found.
 For SN 2000H in which H$\alpha$ is clearly deep ($\tau$$\sim$5 at maximum; Fig. 23), an abundance ratio 
 (i.e. H/He) is estimated to be $\sim$0.23. SN IIb 1993J, around day 16, has a ratio $\sim$0.8. We estimate 
 an even higher ratio for SN 1999em, namely $\sim$3.2. These estimates, although rough, are in accord with 
 our impression that the hydrogen mass 
 increases as we go from Type Ib to IIb to Type II SNe. Figure 32 displays the logarithmic hydrogen to 
 helium abundance ratios against temperature for the events of our sample. Note here that 
 for low temperatures, T $\leq$ 9000, 
 the reported values in Figure 32 can be taken as upper limits. This is because we used a value of 1000 for the 
 ratio $\tau$(H I)/$\tau$(He I). At low temperatures, however, the ratio
 becomes significantly larger than this adopted 
 value (Fig. 2a; Hatano et al. 1999). An additional source of uncertainty may be related
 to the continuum temperature estimates owing to the total reddening effects. 
  Furthermore, according  to our fit experience on the sample spectra, an uncertainty of about 10 $\%$ 
 is assigned to the derived values of the optical depths. In the case of SN IIP 1999em and SN IIb 1993J, with
 their complete H$\alpha$ P-Cygni profiles, greater optical depth uncertainties should be expected, although
 in these events we got good fits to the H$\beta$ profiles. It is worth noting however that the enormous NLTE effects
 in the He I lines lead to very inaccurate determination of helium abundances.

\begin{figure}
\includegraphics[height=8cm,width=9cm]{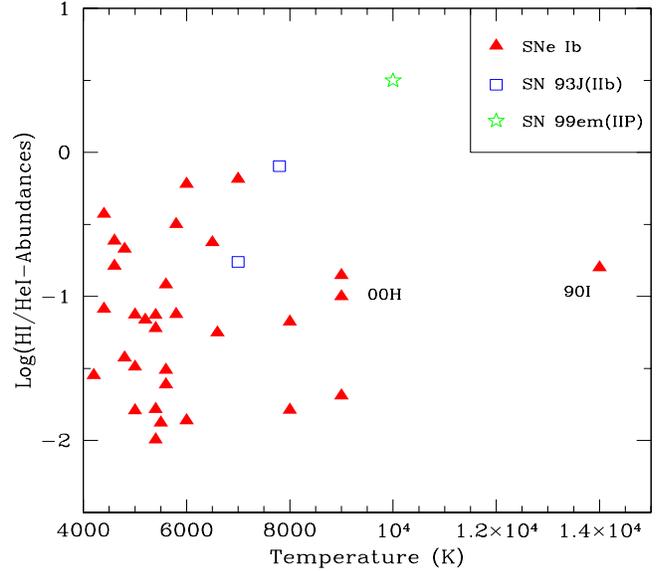}
\caption{ The hydrogen to helium abundances ratio against temperature. See Sect. 4 for more
 details}
\end{figure}
 Making use of the derived photospheric velocities ``$v_{phot}$'', from our synthetic spectra modeling, it is 
 possible to recover ``$spectroscopic$'' estimates of the kinetic energy and mass above the photosphere. In fact, 
 Millard et al. (1999) have shown that for spherical symmetry and an``$r^{-n}$'' density 
 distribution,  the mass (in $M_{\odot}$) and energy (in $10^{51}$ ergs) above the electron-scattering  optical 
 depth `` $\tau_{es}$'' can be expressed as:
\begin{equation}
  M=(1.2 \times 10^{-4})~v_4^{2}~ t_d^{2}~ \mu_e~ \tau_{es}~ f_M(n)
 \end{equation}
\begin{equation}
 E=(1.2 \times 10^{-4})~v_4^{4}~ t_d^{2}~ \mu_e~ \tau_{es}~ f_E(n);
\end{equation}
 where $ f_M(n) = \frac{n-1}{n-3}$ and $ f_E(n) = \frac{n-1}{n-5}$, $t_d$ is time after explosion in days, 
 $v_4$ is $v_{phot}$ in units of 10$^4$ km s$^{-1}$ and $\mu_e$ is the mean molecular weight per free electron.
 
 In applying the above equations to our supernova sample, we adopt $n=8$ and $\tau_{es}=2/3$. For the mean molecular  
 weight per free electron we have different values for each case, namely: SNe Ib: $\mu_e=8$ for half ionized helium 
 or doubly ionized oxygen; SNe Ic: $\mu_e=14$ for a mixture of carbon and oxygen, both singly ionized; 
 SNe II: $\mu_e=1$ to $2$ for fully ionized or half ionized hydrogen. For the case of ``$hybrid$'' IIb SNe we 
 assume a mixture of hydrogen and helium, both half ionized and hence $\mu_e=5$. 

 Figure 33 displays results for five events. Filled symbols refer to the mass (in units of $M_{\odot}$; upper panel), 
 while the open ones indicate the kinetic energy (in units of $10^{51}$ ergs; lower panel). 
 For each supernova, the estimated masses 
 for different epochs are connected by a dotted line. The short-dashed line connects the derived energies 
 in order to clarify trends. For a given phase, the reported amounts indicate the mass moving above the photospheric 
 velocity, carrying a corresponding kinetic energy. The events displayed in Fig. 33 have been selected on the basis 
 of their type and velocity. SN 1990I, being an object with high velocities, appears to have high kinetic energy, 
 while SN 1991D with its lower velocity structure lies at the bottom of the plot. For a significant
  comparison, we choose data at similar phases, around day 38. At that epoch, the mass moving faster than 
 ``$v_{phot}$'' and the corresponding energy are estimated to be: (1.1$~M_{\odot}; 1.6~foe$) for SN 1990I;  
 (0.45$~M_{\odot}; 0.28~foe$) for SN 2000H; (0.3$~M_{\odot}; 0.1~foe$) for SN 1991D; (0.6$~M_{\odot}; 0.8~foe$) 
 for SN 1993J and (0.95$~M_{\odot}; 0.77~foe$) for SN 1994I. SN Ib 1984L 
 has behaviour similar to SN 2000H. These derived quantities suffer, of course, from some uncertainties, namely 
 those incurred with the derived 
 photospheric velocities\footnote{A mean fitting uncertainty of $\sim$500 km s$^{-1}$ is assigned to the derived 
 velocities. The uncertainty in the velocity would affect more 
 the uncertainties in the derived energies than those of the derived masses.} 
from our best fits and the adopted value for the power-index ``$n$''. 
 However, they seem 
 to give reasonable and meaningful values that are not in disagreement with other methods (e.g. light curve modeling
 ; NLTE treatment of early and late spectra), especially 
 for the well studied objects such as SNe 1990I, 1994I and 1993J. The oxygen-mass estimates could be a further 
 check of the trend in the derived photospheric outflow mass. Elmhamdi et al. (2004), on the basis of 
 [O I] 6300,~6364\AA~line analysis at nebular phases, have estimated a lower limit on the oxygen mass to fall in the 
 range  0.7$-$1.35 $M_{\odot}$. The available estimated amounts  for the other objects are as follows: 
 $\sim$0.3 $M_{\odot}$ for SN 1984L (Filippenko et al. 1990\nocite{Fil90});  $\sim$0.5 $M_{\odot}$ for SN 1993J 
 (Houck $\&$ Fransson 1996\nocite{Hou96}) and $\sim$0.4 $M_{\odot}$ for SN 1994I (Woosley et al. 1995\nocite{Woos95}). 
 These estimates follow the trend seen in the masses above the photosphere.  

\begin{figure}
\includegraphics[height=9.5cm,width=9cm]{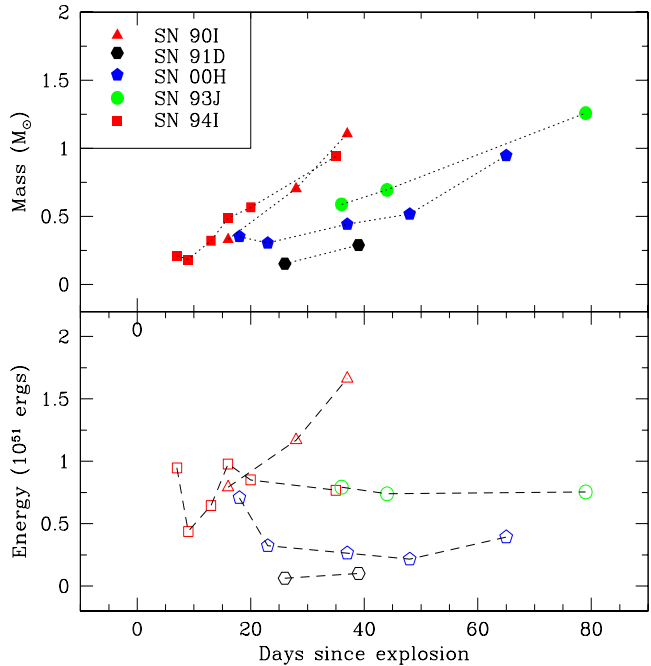}
\caption{ The evolution in time of the ``spectroscopic'' estimates of Mass (upper panel) and  
 Energy (lower panel) above the photosphere.}
\end{figure}
 We emphasize here that this simple approach, using equations 1 and 2, does not provide ``reasonable'' 
 estimates in the case of Type II SNe. In fact when this method is used for both SNe 1987A and 1999em, the 
 estimated ``$M$''  and ``$E$'' above the photosphere seem to be very small, in contradiction with what we may 
 expect. Two possible effects might invalidate the method for Type II objects. First, because the method cannot 
 be used whenever the power index ``$n$'' value falls below 5 (i.e. to keep $f_M(n)$ and $f_E(n)$ positive
 and convergent). For SN 1987A, indeed, Jeffery $\&$ Branch (1990) had found ``$n$''  falling below 5 already within 
 a week after the explosion (see their Figures 15 and 19). Second, in Type II SNe and after the earliest times, 
 the electron-scattering optical depth for the photosphere comes from a thin layer of ionized hydrogen at the 
 recombination front. Farther out, there can be a lot of neutral hydrogen (and helium) that is not accounted for 
 in the ``Millard et al.'' equations adopted in our simple analysis.  When, for instance, ``$\mu_e$'' is set to one,  
 the equations are only giving the mass and energy of the matter near the photosphere. Consequently one may question 
 why they should be taken seriously for the other CCSNe classes. Some carbon and oxygen could be neutral too. One 
 simple explanation could be that because of the large hydrogen
 envelope, these effects seem to be more ``$severe$'' in
  Type II events rather than the rest of SNe. In summary, the simple method does not apply to SNe II because it 
 does not take into account the recombined hydrogen.

 By adopting the estimate of the outflow mass above ``$v_{phot}$'' as an upper
 limit of the helium mass in Type Ib and IIb events, and using the derived constraints on the hydrogen to helium 
 abundances ratio,  one can thus recover an upper and rough estimate of the mass of expelled hydrogen. 
 Around maximum light, 
 the hydrogen amount in SN 1990I is found to be $\sim$0.16$~M_{\odot}$. At similar phase, an amount of about 
 0.1$~M_{\odot}$ is computed for SN 2000H. SN 1993J, however, seems to eject a quantity as high as 
 $\sim$0.7$~M_{\odot}$. SN 1993J indeed serves as a control case for our adopted methodology. Different studies 
 argued that a low-mass hydrogen envelope on a helium core is the most likely scenario for the progenitor of 
 SN 1993J. On the one hand, light curve modeling demonstrates a compatibility fit with a 4$~M_{\odot}$ helium core 
 and hydrogen envelope of $\sim$0.2$~M_{\odot}$ (Woosley et al. 1994)\nocite{Woos94}. 
 A similar analysis invoked a helium core 
 $\sim$4$-6$$~M_{\odot}$, with a residual hydrogen mass less than 0.9$~M_{\odot}$ (Nomoto et al. 1993\nocite{Nom93}; 
 Shigeyama et al. 1994\nocite{Shi94}). On the other hand, best fit models of spectra lead to 
 similar progenitor properties, 
 namely 3.2$~M_{\odot}$ for the helium core with a 0.2$-$0.4$~M_{\odot}$ hydrogen envelope (Houck $\&$ Fransson 1996). 
 Our estimate is therefore in good agreement with the previous cited studies. For the rest of Type Ib objects, an
 upper limit of the hydrogen mass of the order 0.1$~M_{\odot}$ is estimated.

\subsection{Method 2}
 Alternatively, an approximate way to estimate the hydrogen mass invokes 
 the amount needed to fill a uniform density sphere of radius ``$v \times t$'' at an epoch $t_d$ since 
 explosion.
  Adopting a value of ``10$^{-9}$'' for the hydrogen Balmer fraction (within the LTE context), the required 
 ion mass can be given by:
\begin{equation}
 M(M_\odot)\simeq (2.38 \times 10^{-5})~v_4^{3}~ t_d^{2}~\tau(H\alpha);
\end{equation}
  This comes from the equation for the Sobolev optical depth for an expanding envelope (see e.g. Castor
 1970; Jeffery $\&$ Branch 1990).
 where $t_d$ in days and $v_4$ is in 10$^{4}$ km s$^{-1}$. This equation is then used to determine  
 the amount filling a spherical shell of an inner radius ``$v_{min} \times t_d$'' and an outer edge
 of radius ``$v_{max} \times t_d$''. Constraints on the shell widths, ``$v_{min}-v_{max}$'', are 
 approximated using Gaussian fits to the observed H$\alpha$ absorption troughs by means of the corresponding
 FWHM. It is worth noting here that constraints on ``$v_{max}$'' from this method, i.e. Gaussian fit, can be
 considered as a lower limit to the ``real'' maximum velocity.
   
 Although non-thermal excitation and ``NLTE'' effects may be also important for hydrogen, the method seems to 
 give reasonable estimates. For SN 1990I around maximum light a value of 0.02 $M_\odot$ is computed, while at 
 similar phase SNe 1983N and 2000H have, respectively, 0.008 and 0.08 $M_\odot$.
 For type IIb SN 1993J, at day 16 since maximum, the method gives an approximate amount as high as 
 1.7 $M_\odot$. This amount is of course very sensitive to the assumed optical depth. In fact SN 1993J
 has an optical depth similar to Type IIP 1999em (i.e. $\tau(H\alpha)=21$ at $\sim$25days). However
 the fit is more ``convincing'' in SN 1999em than in SN 1993J, especially in the H$\alpha$ absorption
 troughs (see Fig. 30-middle panel and Fig. 31-lower panel). This might be indicative of an overestimated
 optical depth in SN 1993J and/or a larger departure from ``LTE'' compared to SN 1999em.

 A rise time of  $15-20$ days has been adopted in this class of object. Moreover, if we adopt a representative
  H$\alpha$ optical depth of 0.5 at day 20 with an  H$\alpha$ velocity restriction similar to SN 1990I, the
 estimated hydrogen mass is of the order 0.015 $M_\odot$. Events with higher velocity widths and/or deeper H$\alpha$ troughs 
 would eject larger amounts.  
\section{Discussion and conclusion}
 One of our main goals was to identify traces of hydrogen in Type Ib-c, and to identify any systematic similarities 
 and differences correlating with other physical properties. Consequently, guided by modeling early spectra of 
 SN Ib 1990I, 
 we have explored different combinations of ions and shaping the $6000-6500$\AA~wavelength range. Special attention has 
 been devoted to the feature seen near 6300\AA.
 Even though the number of events with spectra well sampled in wavelength, is limited, 
 it appears that hydrogen in varying amounts is identifiable in most Type Ib objects especially at very early times. Only 
 in two cases, namely SNe 1991D and 1991L, does the Ne I 6402\AA~line remain as an alternative possibility with 
 large departure from LTE.
 Even in SN 1996aq, that we classified as a transition Type Ib/c event, and SN 1999ex the presence of the H$\alpha$ 
 trough seems highly preferred.  
\begin{figure}
  \includegraphics[height=9cm,width=9cm]{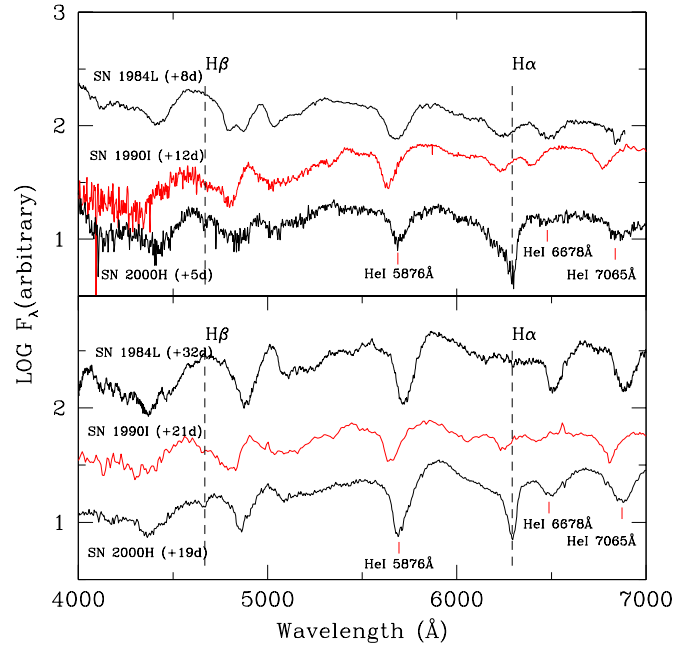}
\caption{Comparison of SN 1990I with SNe 1984L and 2000H, at two different
 phases of evolution. The vertical dashed lines indicate the locations of H$\alpha$ and 
 H$\beta$ troughs in SN 2000H. He I optical series are marked by vertical ticks.}
\end{figure}                   

 At later phases, more than $\sim$ 3 weeks, SNe 1984L, 1988L, 1991ar,
 1998dt, 1998T and 1999dn behave quite similarly, namely still showing evident He I lines 
 and a ``$flat$'' $6000-6500$\AA~region of the spectrum, with a weak 
 H$\alpha$ absorption feature. Even SN 1991D with its particularly low photospheric velocity and narrow lines belongs 
 to this class. SN 1991L appears to be similar to SN 1991D in having shallow He I optical features. SNe 1999di and 2000H 
 are the unique events among the sample that display obvious He I lines, together with a pronounced and deep 
 H$\alpha$ absorption line. It is interesting to note here the peculiar behaviour of SN 1997dc. The event has He I 
 troughs as deep as in SNe 1999di and 2000H, however it has a flat $6000-6500$\AA~region. 
  SN 1990I at this phase shows 
 distinct He I lines and a moderate H$\alpha$ trough. Figure 34 compares SN 1990I with SNe 1984L and 2000H 
 at two different phases of evolution. The vertical 
 dashed lines indicate the locations of H$\alpha$ and  H$\beta$ troughs in SN 2000H. On the one hand, while at early 
 phases (upper panel) the trough assigned to H$\alpha$ is clear in all the three
 SNe, it disappears later on in SN 1984L (lower panel). On the other
 hand, H$\beta$ appears clearly in SN 2000H and absent 
 in SNe 1984L and 1990I. Moreover, the comparison emphasizes the high velocity behaviour of 
 SN 1990I as is evident from the blueshifted features in SN 1990I compared to the others.
  He I optical lines are also shown. For the three events the He I troughs at 5876\AA, and 7065\AA~are 
 clearly identified and get narrower with time. At early phase the He I 6678\AA~is recognized in SNe 1990I and 1984L
 while it is just hinted in SN 2000H. Later on, Fig.34-bottom panel, the  He I 6678\AA~line grows in strength in
 SNe 1984L and 2000H while it fades in SN 1990I. The comparison of these three events, in terms of changes in line
  visibility and velocity, demonstrates the complexity 
 involved in demonstrating any thing like a continuous sequence. 

\begin{figure}
\includegraphics[height=10.5cm,width=9cm]{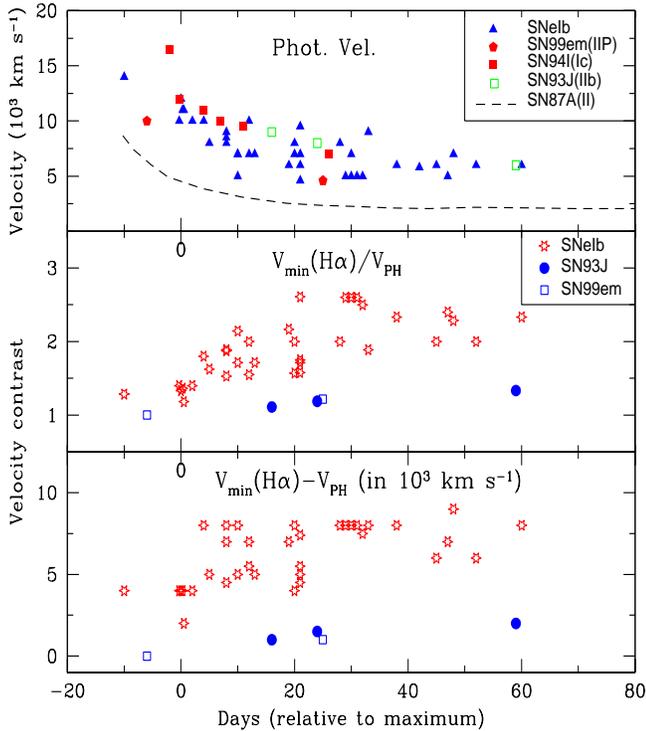}
\caption{The photospheric velocity evolution of Ib SNe sample, compared to SNe 1987A, IIP 1999em and Ic 1994I 
 (upper panel). Middle panel: the H$\alpha$ contrast velocity ratio evolution 
  (i.e. v$_{min}^{ratio}$(H$\alpha$)$/$v$_{phot}$). Lower panel: the H$\alpha$ contrast velocity evolution 
 (i.e. v$_{min}$(H$\alpha$)-v$_{phot}$). Results for SN IIP 1999em and SN IIb 1993J are also shown.}
\end{figure}

 Hydrogen manifests its presence in a different way in the other CCSNe types. A part from SN 1996aq that we re-classify 
 as a transient Ib/c event rather than a pure Type Ic SN, there is no evidence for the presence of hydrogen in Type Ic 
 objects analyzed here (i.e. SNe 1987M and 1994I). The ``$hybrid$'' SN IIb 1993J displays typical Type II features at 
 early phases, such as strong and broad H$\alpha$ P-Cygni emission component, which is absent in Type Ib objects. This 
 can be explained within the context of the ``detachment'' concept. In fact, H$\alpha$ P-Cygni profile would lose its obvious 
 emission component when it is highly detached.
 At early epochs, both SN IIb 1993J and SN II 1999em show clear evidence of the other H I Balmer lines since the 
 corresponding optical depths are too large to allow them to be distinctly visible, contrary to normal Type Ib  
 objects, with the exception of some cases with deep and conspicuous H$\alpha$ troughs that present signature of 
 H$\beta$ too (e.g. SNe 1999di and 2000H). Two factors
 make H$\beta$ barely discernible in Type Ib: the optical depth found to fit H$\alpha$ is small and the contrast velocity 
 of H$\alpha$ is high. The opposite is seen in Type II and IIb  events.
 As time goes on, the H$\alpha$ emission peak in SN 1993J changes to a 
 double-peaked structure which we recognized as the emergence of the He I 6678\AA, with more clearer He I optical lines. 
 These laters disappear at the nebular phase and the spectrum is
 dominated by typical Ib features such as
 [O I], [Ca II] and Ca II in emission. The situation in SN~1999em, and in Type II  in general, is different. The 
 He I lines are found to disappear very early, about one weak after explosion, and even at later nebular epochs 
 the H$\alpha$ is still the most prominent feature in the spectrum.

 In Figure 35, upper panel, we report the resulting photospheric velocities from our best fits for the CCSNe sample. Data 
 for SN 1987A, corresponding to Fe II 5018\AA~absorption, are also shown for comparison (dashed line;  Phillips 
 et al. 1988\nocite{Phil88}). Additional points for SN 1994I are taken from Millard et al. 1999. The plot 
 indicates the low velocity 
 behaviour of Type II SNe, both 1987A and 1999em, at early as well as intermediate epochs. SN IIb 1993J follows somewhat 
 similar behaviour as SN Ic 1994I in having higher velocities. 
 As far as Type Ib SNe are concerned, they appear to display a different velocity evolution. The scatter seems to increase at 
 intermediate  phases (around 20$-$30 days). This fact can be simply due to the paucity of available observations 
 outside that range. Around day 20, for example, a scatter as high as 5000 km s$^{-1}$ is measured. SNe 1990I and 1998dt 
 belong to a class with the higher ``$v_{phot}$'', while objects such as SNe 1991D and 1996aq have the lowest estimated 
 velocities, approaching even Type II objects. The remainder of the Ib events follow a similar trend, namely the one 
 described by Branch et al. (2002). SN 1999ex is found to belong to this class. Surely more data are needed to obtain 
 meaningful statitical conclusions.
\begin{figure}
\includegraphics[height=9cm,width=9cm]{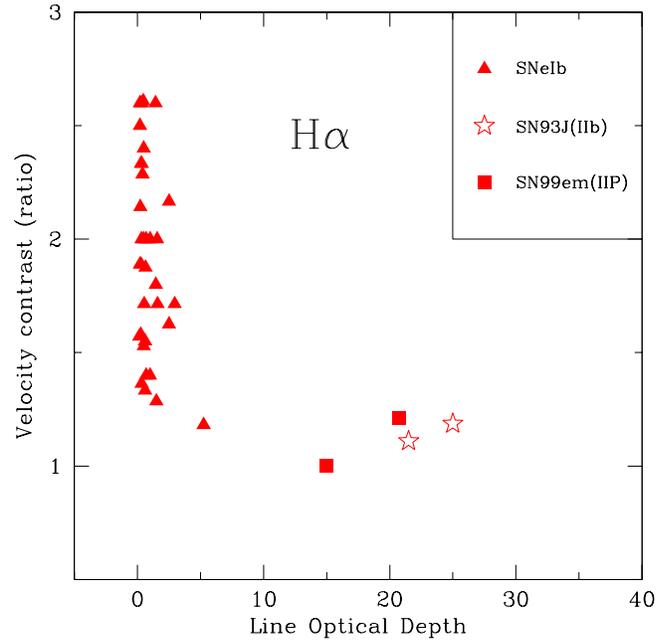}
\caption{ The H$\alpha$ velocity contrast evolution, $v_{cont}^{ratio}$(H$\alpha$), versus H$\alpha$ optical depth. 
 Results for SNe IIP 1999em and IIb 1993J are displayed for comparison. Note that results are displayed independently
 of the phase.}
\end{figure}
  
  The middle and bottom panels in Fig. 35 display the evolution the of H$\alpha$ contrast velocities,
  ``$v_{cont}^{ratio}$'' and ``$v_{cont}$'' respectively,  for the 
 sample. As discussed before, when discussing differences in the H$\alpha$ profile in CCSNe, we would expect an 
 increasing value of ``$v_{cont}$(H$\alpha$)''  going from Type II to IIb to Ib SNe. This trend is in fact illustrated 
 in the plot (bottom panel). A similar trend is seen in the ``$v_{cont}^{ratio}$(H$\alpha$)'' evolution
 as indicated by the middle panel. While in SN IIb 1993J and 
 SN~IIP~1999em the line is found
 to be either undetached or slightly detached, it is 
 highly detached in Type Ib events. Moreover, the ``$v_{cont}$(H$\alpha$)'' is found to increase within the first 15 days, 
 reaching values as high as 8000  km s$^{-1}$, and then follows an almost constant evolution. According to Table 2 
 and up 
 to $\sim$60 days, Type Ib SNe have hydrogen down to 11000-12000 km s$^{-1}$, while in SN 1993J hydrogen is down to 
 8000 km s$^{-1}$. SNe II appear to have hydrogen down to even lower velocities ($\sim$5000 km s$^{-1}$ in 
 SN 1999em).
 In addition, hydrogen in Type Ib SNe is found to have very small optical depths 
 independently of the contrast 
 velocity (also independently of the phase). This is shown by Figure 36. In fact, Type Ib objects populate the
  region of extremely low optical depths, 
 although the ``$v_{cont}^{ratio}$(H$\alpha$)'' spans a large range, contrary to SNe 1993J and 1999em.
\begin{figure}
\includegraphics[height=9cm,width=9cm]{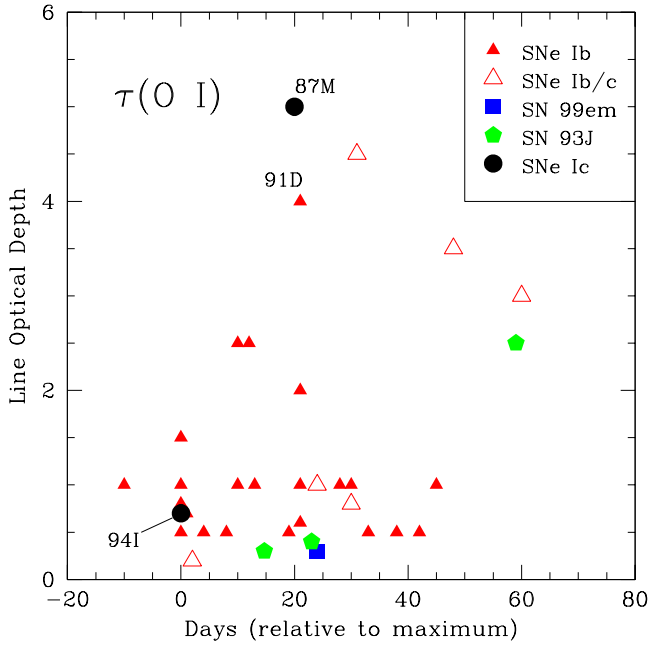}
\caption{ The optical depth of the O I 7773\AA~line according to the best fit spectra. Data for 
 different SN types are reported for comparison.}
\end{figure}

 Table 2 also displays properties of the helium line. In Type Ib SNe, He I lines are found to be not always 
 detached as is the case for hydrogen. The He I 5876\AA~line is found to be clearly distinguishable 
 in Type Ib objects. A contribution 
 from Na ID is argued for in some cases, such as the broad feature in SN 1996aq and cases with an observed bump 
 shortward of the minimum absorption as seen in SNe 1990B and 1997dc. It is important to recall here that in some cases
 we obtained a better fit to the He I optical lines adopting an $e-$folding SSp, which allows a two-component 
 treatment of the line and a gradually decreasing optical depth below the detachment velocity, rather than a 
 discontinuity in it. In the case of SN 1999ex, we extend our He I line fits beyond 1 micron,
 confirming the presence of various IR lines, namely at 1.083, 1.284, 1.700(?), and 2.085 microns (Fig. 20),
  in accord with optical He I parameters.
 In our analysis, we also pointed to some interesting line identifications.  For example the 
 contribution of Sc II lines in nearly all the CCSNe spectra, especially for reproducing the double absorption features 
 blueward of the He I 5876\AA~line. We note here that except for He I and H I, almost all the 
 lines in the CCSNe synthetic spectra are undetached.  

 Another important study is the behaviour of the O I 7773\AA~line. Figure 38 reports the resulting optical depths 
 corresponding to our best synthetic spectra fits. It is not easy to draw clear-cut conclusions from this figure since 
 one needs, for instance, to populate the figure with more Type Ic objects. However, at intermediate phases, it seems 
 that Type Ib objects tend to concentrate in the low optical depth region, while SN Ic 1987M is found to display the 
 deepest profile. SNe 1993J and 1999em, at similar phases, are the objects with the lowest O I 7773\AA~optical depth. 
 At somewhat later epochs, transient Type Ib/c objects display deep O I 7773\AA~troughs. The stronger and deeper 
 permitted oxygen lines at early phases of SNe Ic and Ib/c spectra might imply that they are less diluted 
 by the presence of a helium envelope. Indeed one might expect oxygen lines to be more prominent for a
 ``naked'' C/O progenitor core. Despite the paucity of well sampled CCSNe observations, two observational
 aspects tend to reinforce this belief: $\bf{First}$, the forbidden
 lines, especially [OI]6300,~6364\AA, seem to appear
 earlier following a SNe sequence ``Ic$-$Ib$-$IIb$-$II''. In fact the oxygen line emerges at an age of 1-2 months in
 Type Ic SN 1987M (Filippenko 1997)\nocite{Fil97}. SN Ic 1994I displayed evidence for the line at an age
 of 50 days, although some hints may even be seen in the $\sim$36day spectrum (Clocchiatti et al. 1996b). While
 in SN Ib 1990I it was hinted at the 70day spectrum (Elmhamdi et al. 2004). In other Type Ib SNe it appears
 earlier than in SN 1990I. In SN IIb 1993J, a transition object, the line was visible in the 62day
 spectrum (Barbon et al. 1995)\nocite{Barb95}. SN 1996cb, another well observed IIb event, showed
 evidence of the [OI]6300,~6364\AA~line around day 80 (Qiu et al. 1999).
 In SNe II, however, the line appears later: around day 150 in SN 1987A (Catchpole et al. 1988)\nocite{Cat88}
 and after day 138 in SN 1992H (Clocchiatti et al. 1996a). In SN II
 1999em it is suggested at a somewhat
 earlier phase compared to SNe 1987A and 1992H, namely at day 114. Whether all this can be understood in terms
 of the lower progenitor mass of SN 1999em and its presumed lower oxygen mass remains to be
 investigated (Elmhamdi et al. 2003). $\bf{Second}$, it seems that the nebular emission line decreases
 in breadth following the SNe sequence above. Of course much work is
 still needed in this respect based on larger CCSNe samples at both photospheric and nebular phases, relating them to the
 photometry of the CCSNe variety (Elmhamdi $\&$ Danziger, in preparation). Furthermore, one important
 point is to check for any possible correlations between the oxygen minimum velocity (at early epochs),
 their line widths, and to relate this to a progenitor mass indicator [Ca II]/[O I] ratio (at nebular
 phases; Fransson $\&$ Chevalier 1989), the progenitor properties (i.e. masses, energies). Other factors
 such as mixing, variation in envelope densities and metallicity may play an additional complicating role.

 Finally we presented two methods to determine the ejecta and hydrogen mass in CCSNe and 
 especially in Type Ib events. Although the methods are very approximate\footnote{``Method 2'' is a direct method, 
 while ``Method 1'' is undirect and gives upper limits on the hydrogen mass.}, our results do not conflict 
 with more detailed estimates, those based 
 on hydrodynamical and NLTE models. A thin layer of hydrogen, ejected at high velocities down to 
 11000-12000 km s$^{-1}$, appears to be present in almost all the Type Ib events studied here. These results suggest 
 possible directions for further more sophisticated work. 
 The necessity to introduce lines of Ne I, Sc II, Ba II, Ca I, Fe II in some cases but not
 in others must inevitably raise concerns about the identifications and other reasons for these variations. 
\begin{acknowledgements}
 A. Elmhamdi is grateful to the ESO (European Southern Observatory) support that allowed his stay
 at Garching where part of the work has been done. I.J. Danziger was supported in this work by MIUR with a grant 
 for PRIN 2004 from the Italian Ministry of Education. Research on supernovae at Harvard University 
 is supported by NSF Grant AST-0205808. Supernova research at Oklahoma University is supported by NSF grants 
 AST-0204771 and AST-0506028, and NASA LTSA grant NNG04GD36G.  

\end{acknowledgements}

\begin{table*}
\begin{minipage}{100mm}
\setlength{\tabcolsep}{3pt}
 \caption{Main data and fitting parameters of the CCSNe sample spectra} 
\bigskip
\centering
\begin{tabular}{cccccccccccc}
\hline \hline
Supernova  & Host Galaxy & $V_{recession}$ & Phases & $V_{phot}$ & $T_{bb}$$^{1}$&$\tau(H)$& $\tau(He I)$&$V_{min}(H)$&$V_{min}(He I)$& Nbr(ions) \\
(Type) &  & ($km s^{-1}$) & (days) & ($km s^{-1}$) &($K$)&&& ($km s^{-1}$) &  ($km s^{-1}$) & \\
\hline  \hline
SN1990I (Ib) &NGC 4650A &2902 &max&12000&14000&0.6&2.9&16000&14000&7   \\
 && &12&10000&5500&0.6&1.85&15500&13000&7   \\
 && &21&9500&5400&0.26&0.94&15000&12000&7   \\
\hline 
SN1983N (Ib) &NGC 5236 &513 &max&11000&8000&0.34&2.5&15000&$V_{phot}$&8   \\
 && &10&7000&5000&0.55&3.4&12000&9000&11   \\
\hline 
SN1984L (Ib) &NGC 991 &1534 &8&8000&8000&0.65&2.5&15000&10000&5   \\
 && &32&5000&5600&0.2&6.7&12500&7000& 5 \\
\hline 
SN1987M$^{2}$ (Ic) &NGC 2715 &1339 &20&7000&4500&$----$&0.6&$----$&9000&7   \\
\hline 
SN1988L (Ib/c) &NGC 5480 &1856 &20&8000&9000&0.7&1&16000&11000&9   \\
\hline 
SN1990B (Ib/c) &NGC 4568 &2320 &8&8500&5400&0.5&0.55&13000&10000&7   \\
 && &30&7000&4000&$----$&0.8&$----$&10000& 9 \\
\hline 
SN1991ar (Ib) &IC 49 &4562 &28&8000&4400&0.32&10&16000&$V_{phot}$&10   \\
 && &48&7000&4800&0.4&4.8&16000&8000&10  \\
 && &60&6000&4600&0.28&4.4&14000&7000&12 \\
\hline 
SN1991D (Ib) &PGC 84044 &12500 &21&4600&7000&0.46&1.8&12000&6000&11   \\
\hline 
SN1991L (Ib/c) & M 7-34-134 &9050 &31&5000&6000&0.34&2&13000&$V_{phot}$&12  \\
\hline 
SN1993J (IIb) &M 81 &35 &16&9000&7800&21&0.6&10000&$V_{phot}$&10  \\
 && &24&8000&7000&25&1.5&9500&$V_{phot}$&10  \\
 && &59&6000&5000&2&30&8000&$V_{phot}$&11 \\
\hline 
SN1994I (Ic) &NGC 5194 &461 &max&12000&7000&$----$&$----$&$----$&$----$&9  \\
\hline 
SN1996aq (Ib/c) &NGC 5584 &1602 &2&9000&9000&0.12&0.35&15000&12000&8  \\
&& &24&4600&5000&$----$&0.25&$----$&7000& 11 \\
\hline 
SN1997dc (Ib) &NGC 7678 &3480 &28&5000&4200&0.2 &10 &13000&7000& 10 \\
\hline 
SN1998dt (Ib) &NGC 945 &4580 &8&9000&5600&0.3 &4 &17000&11000& 6 \\
&& &33&9000&5000&0.2&10&17000&$V_{phot}$&9  \\
\hline 
SN1998T (Ib) &NGC 3690 &3080 &21&6000 &5400&0.34&1.15 &10500 &9000 &9 \\
&& &42&5800&5200&$----$&0.75&$----$&8000&9  \\
\hline 
SN1999di (Ib) &NGC 776 &4920 &21&7000&4800&3 &12.6 &12000&8500& 8 \\
&& &45&6000&5200&1.6&10&12000&7500&9  \\
&& &52&6000&5000&1&10&12000&7000&9  \\
\hline 
SN1999dn (Ib) &NGC 7714 &4920 &-10&14000&6600&1.5 &2 &18000&$V_{phot}$& 5 \\
&& &max&10000&6000&1&5&14000&11000&6  \\
&& &10&7000&5800&0.24&1.9&15000&9000&10  \\
&& &38&6000&5400&0.34&14.5&14000&7000&10 \\
\hline 
SN1999ex (Ib) & IC 5179  &3498 &4&10000&5800&1.45 &2.35 &18000&11000& 7 \\
&& &13&7000&5600&1.6&3.4&12000&8000&12 \\
\hline 
SN1999em (IIP) & NGC 1637&717 &-6&10000&10000&15 &0.6 &$V_{phot}$&$V_{phot}$& 3 \\
&& &25&4600&6400&21&$----$&5600&$----$&13 \\
\hline 
SN2000H (Ib) & IC 454&3894 &max&11000&9000&5.25 &2 &13000&$V_{phot}$& 9 \\
&& &5&8000&6500&2.5&2&13000&9000&9 \\
&& &19&6000&4600&2.5&5&13000&8000&9 \\
&& &30&5000&6000&1.45&3.4&13000&7000&11 \\
&& &47&5000&4400&0.5&1&12000&7000&9 \\
\hline 
\end{tabular}
\end{minipage}\\
\\
{\emph{{\rm \scriptsize
$1$- The events are corrected for the galactic reddening effect (Schelegel et al. 1998). Additional 
 host galaxy\\ $~~~~~$extinction is corrected for only in two objects, namely SNe 1990I and 1999em.\\
$2$- The reported data corresponds to the low He I velocity case (see Sect. 3.2)
}}}\\ 
\normalsize
\end{table*}

\end{document}